\documentclass[preprint,12pt]{elsarticle}
\biboptions{numbers,sort&compress}
\newcommand{\reffig}[1]{Fig. \ref{#1}}
\newcommand{\reftab}[1]{Table \ref{#1}}
\newcommand{\refequ}[1]{(\ref{#1})}
\newcommand{\tabincell}[2]{\begin{tabular}{@{}#1@{}}#2\end{tabular}}
\DeclareMathAlphabet{\mathscr}{OT1}{pzc}{m}{it}

\usepackage{threeparttable}
\usepackage{graphicx}
\usepackage{amssymb}
\usepackage{upgreek}
\usepackage{siunitx}
\usepackage{subfigure}
\usepackage{color, xcolor}
\usepackage{caption}

\usepackage{lineno}
\usepackage{amsmath}
\usepackage{threeparttable}
\usepackage{multirow}
\usepackage{marvosym}
\usepackage{makecell}
\usepackage{nomencl} 
\usepackage{ifthen}
\usepackage{soul}
\usepackage{adjustbox}
\renewcommand{\nomgroup}[1]{%
	\item[\textbf{%
		\ifthenelse{\equal{#1}{P}}{Parameters and variables}{}%
		\ifthenelse{\equal{#1}{S}}{Superscripts and subscripts}{}
	}]%
}
\makenomenclature

\journal{Applied Energy}

\makeatletter
\def\@author#1{\g@addto@macro\elsauthors{\normalsize%
		\def\baselinestretch{1}%
		\upshape\authorsep#1\unskip\textsuperscript{%
			\ifx\@fnmark\@empty\else\unskip\sep\@fnmark\let\sep=,\fi
			\ifx\@corref\@empty\else\unskip\sep\@corref\let\sep=,\fi
		}%
		\def\authorsep{\unskip,\space}%
		\global\let\@fnmark\@empty
		\global\let\@corref\@empty  
		\global\let\sep\@empty}%
	\@eadauthor={#1}
}
\makeatother

\begin{document}
\captionsetup[figure]{name={Fig.},labelsep=period}
\begin{frontmatter}


\title{Thermal Modelling and Controller Design of an Alkaline Electrolysis System under Dynamic Operating Conditions}


\author[label1]{Ruomei Qi \corref{cor0}}
\author[label1]{Jiarong Li \corref{cor0}}
\author[label1,label2]{Jin Lin \corref{cor1}}
\cortext[cor0]{These authors contributed equally.}
\cortext[cor1]{Corresponding author}
\ead{linjin@tsinghua.edu.cn}
\author[label1,label3]{Yonghua Song}
\author[label4,label5]{Jiepeng Wang}
\author[label5]{Qiangqiang Cui}
\author[label6]{Yiwei Qiu}
\author[label2]{Ming Tang}
\author[label2]{Jian Wang}
\address[label1]{State Key Laboratory of Control and Simulation of Power Systems and Generation
Equipment, Department of Electrical Engineering, Tsinghua University, Beijing, China}
\address[label2]{Tsinghua-Sichuan Energy Internet Research Institute, Chengdu, China}
\address[label3]{Department of Electrical and Computer Engineering, University of Macau, Macau, China}
\address[label4]{School of Materials Science and Engineering, Shanghai University, Shanghai, China}
\address[label5]{Purification Equipment Research Institute of CSIC, Handan, China}
\address[label6]{College of Electrical Engineering, Sichuan University, Chengdu, China}

\begin{abstract}
Thermal management is vital for the efficient and safe operation of alkaline electrolysis systems. Traditional alkaline electrolysis systems use simple proportional-integral-differentiation (PID) controllers to maintain the stack temperature near the rated value. However, in renewable-to-hydrogen scenarios, the stack temperature is disturbed by load fluctuations, and the temperature overshoot phenomenon occurs which can exceed the upper limit and harm the stack. This paper focuses on the thermal modelling and controller design of an alkaline electrolysis system under dynamic operating conditions. A control-oriented thermal model is established in the form of a third-order time-delay process, which is used for simulation and controller design. Based on this model, we propose two novel controllers to reduce temperature overshoot: one is a current feed-forward PID controller (PID-I), the other is a model predictive controller (MPC). Their performances are tested on a lab-scale system and the experimental results are satisfying: the temperature overshoot is reduced by 2.2$^{\circ}$C with the PID-I controller, and no obvious overshoot is observed with the MPC controller. Furthermore, the thermal dynamic performance of an MW-scale alkaline electrolysis system is analyzed by simulation, which shows that the temperature overshoot phenomenon is more general in large systems. The proposed method allows for higher temperature set points which can improve system efficiency by 1\%.

\end{abstract}


\end{frontmatter}
\nomenclature[P]{$P$}{Electricity power}%
\nomenclature[P]{$Q$}{Thermal power}%
\nomenclature[P]{$U$}{Voltage}%
\nomenclature[P]{$I$}{Current}%
\nomenclature[P]{$T$}{Temperature}%
\nomenclature[P]{$\bar{T}$}{Average temperature}%
\nomenclature[P]{$C$}{Thermal capacity}%
\nomenclature[P]{$t$}{Time}%
\nomenclature[P]{$\tau$}{Time-delay}%
\nomenclature[P]{$c$}{Specific heat capacity}%
\nomenclature[P]{$v$}{Volume flow rate}%
\nomenclature[P]{$\rho$}{Density}%
\nomenclature[P]{$k$}{Heat transfer coefficient}%
\nomenclature[P]{$A$}{Area}%
\nomenclature[P]{$R$}{Thermal resistance}%
\nomenclature[P]{$y$}{Valve opening}%
\nomenclature[P]{$\hat{y}$}{Control signal for valve opening}%
\nomenclature[P]{$y$}{Valve opening}%
\nomenclature[P]{$\eta$}{Electrolysis efficiency (HHV)}%
\nomenclature[P]{$\eta_I$}{Current efficiency}

\nomenclature[S]{*}{Steady-state}%
\nomenclature[S]{ele}{Electrolysis}%
\nomenclature[S]{dis}{Heat dissipation}%
\nomenclature[S]{th}{Thermal neutral}%
\nomenclature[S]{sep}{Separator}%
\nomenclature[S]{c}{Cooling water}%
\nomenclature[S]{amb}{Ambient}%

\printnomenclature

\section{Introduction}
\label{S:Introduction}
Thermal management is one of the most important auxiliary units in alkaline electrolysis systems \cite{Thermal management-1,Thermal management-2}. Since the electrolysis reaction is exothermic at room temperature, cooling devices are equipped to maintain the system temperature at the rated temperature \cite{Japan}. Temperature affects both system efficiency and security: temperatures lower than the rated temperature hinder the electrolysis reaction and lead to low efficiency \cite{Thermodynamic}; on the other hand, high temperatures beyond the upper limit can harm and shorten the lifetime of the materials of the stack by decreasing the corrosion resistance \cite{book,Claushthal-2021,3MW}. A consequence of corrosion is that the diffusion of gas molecules through the diaphragm increases, negatively impacting the gas purity \cite{3MW,2008Thermal}. To prevent temperature deviation from the rated temperature, a PID temperature controller is used, which suppresses the influence of external disturbance by regulating the cooling water flow rate \cite{3MW}. 

However, the dynamic operation mode becomes routine in renewable to hydrogen scenarios, in which an alkaline electrolysis system converts surplus renewable electricity into hydrogen \cite{Renewable,Renewable2}. The system temperature is greatly disturbed by load fluctuations, which cannot be fully suppressed by a traditional PID temperature controller \cite{Temperature fluctuation,3MW}. Particularly, when the load suddenly increases, the stack temperature overshoots beyond the rated value due to increased heat production, and the large inertia as well as delay in the heat transfer process hinders the immediate functioning of the cooling system. To avoid temperature overshoot damaging the stack, one engineering method is to adopt a lower temperature set point; however, this sacrifices the electrolysis efficiency. Another method is to adopt a smaller load ramping rate, which results in poor dynamic operation performance.

Thermally related models are needed to analyze the temperature fluctuation under dynamic operation, which includes two parts: the thermodynamic model and the thermal model. The thermodynamic model illustrates the influence of temperature on system consumption and efficiency. For the electrolysis cell, the cell voltage, as well as the cell consumption, decreases with increasing temperature due to the beneficial effect of a high temperature environment on the thermodynamics of the electrolysis reaction \cite{M-S-mechanism-1}. The relationship between cell voltage and temperature can be described by either a mechanism model \cite{M-S-mechanism-1,M-S-mechanism-2,M-S-mechanism-3,M-S-mechanism-4,M-S-mechanism-5-RC} or an empirical model \cite{M-SD-experience-1-Ulleberg,M-S-experience-2-flowrate, M-S-experience-3-15kW,M-SD-experience-4,M-S-experience-5-Resistance,M-S-experience-6}. For auxiliary devices, the consumption of the hydrogen compressor, cooling fan, electrolyte circulation pump and heater also depends on temperature \cite{Temperature effect}. Jang et al. \cite{Temperature effect} developed an AEL system model and simulated the system consumption with various operating temperatures. The results show that in the high current density region, a higher operating temperature increases the system efficiency; in the low current density region, the optimal temperature depends on the heater's consumption.

Thermal models are used to predict the system temperature based on the analytical characterization of thermal accumulation, production and exchange \cite{Review}. According to the modelling complexity, existing thermal models can be classified into first-order lumped models \cite{M-SD-experience-1-Ulleberg,2008Thermal,Japan,M-S-experience-6,M-S-mechanism-1,Dispatch,CiteUlleberg} and multi-order models \cite{3MW,Design consideration}. The most well-known is the lumped model developed by Ulleberg \cite{M-SD-experience-1-Ulleberg}, in which the overall thermal energy balance considers heat generation, the heat loss to the ambient and the  auxiliary cooling demand. This model \cite{M-SD-experience-1-Ulleberg} is widely used in studies \cite{Japan,M-S-experience-6,M-S-mechanism-1,Dispatch,CiteUlleberg}. Other first-order models are modified based on \cite{M-SD-experience-1-Ulleberg}. Dieguez et al. \cite{2008Thermal} included an extra term in the thermal energy balance equation that accounts for the sensible and latent heat removed with hydrogen and oxygen streams leaving the system, as well as the sensible heat required to warm deionized water from room temperature to the stack operating temperature. First-order models can only predict the average temperature of the system, and it is difficult to consider multiple temperature nodes, such as the before-stack and after-stack temperatures. Indeed, the after-stack temperature is generally the hottest in the system and should be specially considered. Sakas et al. \cite{3MW} established a second-order thermal model including two thermal capacitances: the stack and the gas-liquid separator; however, the model is only verified by the experimental results of single temperature measurement point and doesn't show the characteristics of a multi-node model. Rizwan et al. \cite{Design consideration} adopted a third-order thermal model, in which the energy balance equations for the stack and the hot and cold sides of the heat exchanger are written separately; however this model has not been verified, and only the steady-state form is used in the simulation. Furthermore, none of the abovementioned thermal modelling studies consider the time-delay caused by the heat medium convection, e.g. electrolyte and cooling water circulations, which is significant for the accurate prediction of thermal dynamic processes, including temperature overshoot and oscillation. In other words, there is a lack of thermal modelling research from a control perspective.

Research on temperature control is also limited. Only Sakas et al. \cite{3MW} considered a PID temperature controller in the simulation; however, the dynamic performance of the temperature controller was not discussed. 

This paper focuses on thermal modelling and controller design under dynamic operating conditions. Novel temperature controllers are designed based on the idea of feed-forward to mitigate temperature fluctuations and improve system efficiency as well as dynamic performance. The main contributions are as follows:
\begin{enumerate}
	\item A control-oriented thermal model for alkaline electrolysis system is established with the form of a third-order time-delay process. Its uniqueness lies in its ability to predict both  before-stack and after-temperatures rather than the system's average temperature; therefore it is suitable for temperature controller design. Furthermore, the introduction of time-delay terms helps to describe the thermal process (such as overshoot and oscillation) more accurately.  

	\item To mitigate temperature fluctuation under dynamic operating conditions, two novel temperature controllers are designed, including a current feed-forward PID controller, which is easy to use, and a more complicated MPC controller which is suitable for scenarios that can obtain future load information in advance, e.g. peak shaving scenario. Both controllers help reduce the temperature overshoot and make higher temperature set points possible to improve the system efficiency.
	
\end{enumerate}

The paper is organized as follows. Section \uppercase\expandafter{\romannumeral2} introduces the basic concepts of thermal management in alkaline electrolysis systems. Section \uppercase\expandafter{\romannumeral3} derives the thermal model. Section \uppercase\expandafter{\romannumeral4} proposes a design method for temperature controllers. Section \uppercase\expandafter{\romannumeral5} verifies the thermal model by experimental results. Section \uppercase\expandafter{\romannumeral6} tests the proposed temperature controller on a laboratory-scale test platform. Section \uppercase\expandafter{\romannumeral7} shows the simulation results for a MW-scale alkaline electrolysis systems.

\section{Thermal management of an alkaline electrolysis system}
The basic concepts of thermal management in alkaline electrolysis systems are introduced in this section. Based on the thermal management process, three operation regions are divided according to different thermal dynamic characteristics. The trade-off between safety and efficiency in high-load region is highlighted which shows the importance of temperature control.
\subsection{Thermal management method}
\label{S:System process}
\begin{figure}[htb]  
	\makebox[\textwidth][c]{\includegraphics[width=1.4\textwidth]{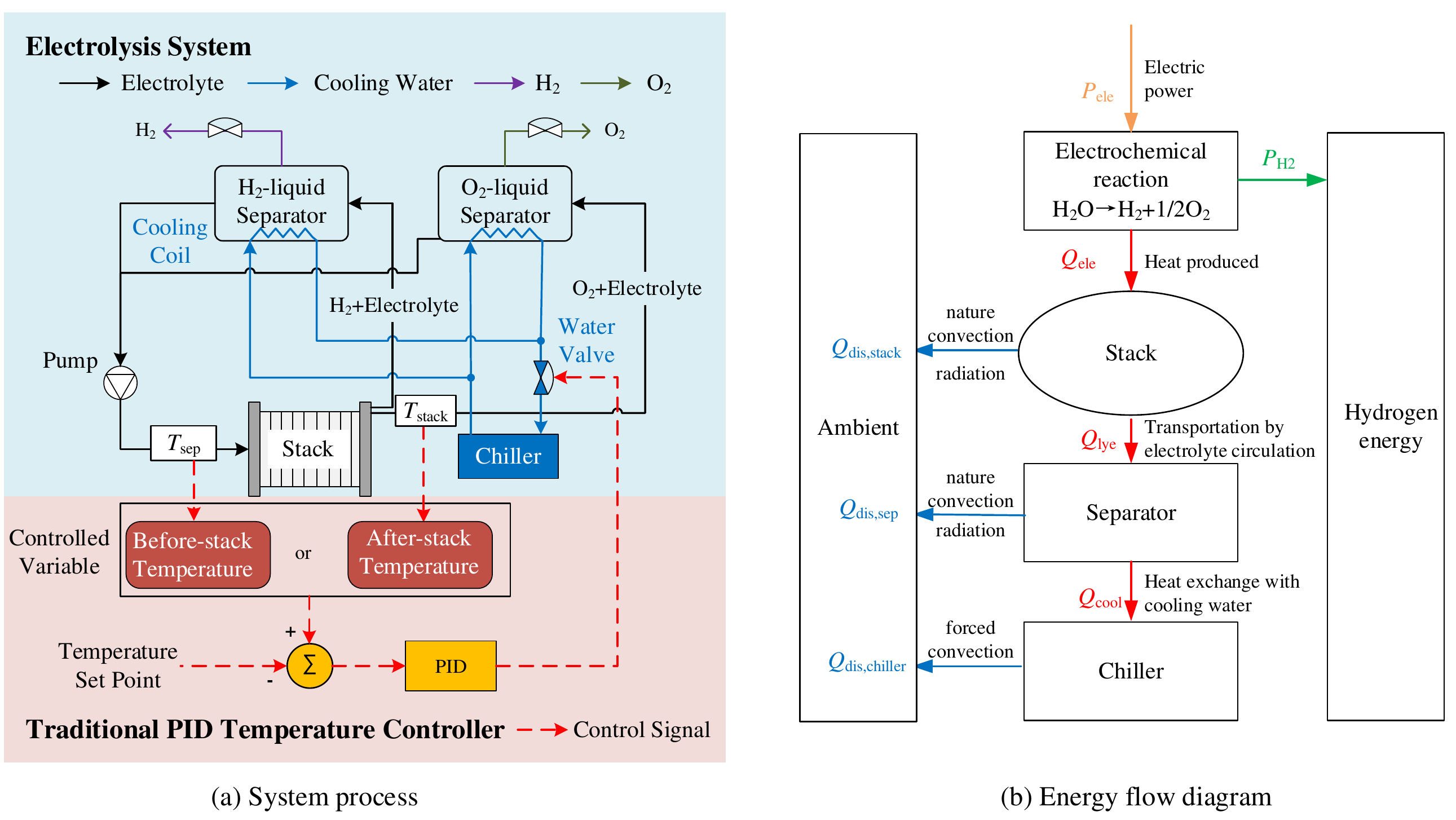}}  
	\caption{Thermal management of an alkaline electrolysis system.}
	\label{Fig1System}
\end{figure}

The process of the alkaline electrolysis system studied in this paper is shown in \reffig{Fig1System}(a), which has the same structure as \cite{Pressure control,Huaneng}. The corresponding energy flow diagram is shown in \reffig{Fig1System}(b).

The stack is the core element of the system, in which water is electrolyzed to produce hydrogen and oxygen products. The electrolysis reaction is exothermic at room temperature. When electric power input is converted to hydrogen energy, heat is produced due to the energy loss in the reaction, which in turn increases the stack’s temperature \cite{Japan}. The relationship among electric power input $P_\mathrm{ele}$, heat produced $Q_\mathrm{ele}$ and hydrogen energy $P_\mathrm{H_2}$ is as follows:
\begin{subequations}
	\begin{equation}
	P_\mathrm{ele}=U_\mathrm{cell}I_\mathrm{cell}N_\mathrm{cell}
	\end{equation}
	\begin{equation}\label{eq:Heatproduced}
	Q_\mathrm{ele}=(U_\mathrm{cell}-U_\mathrm{th})\eta_I I_\mathrm{cell}N_\mathrm{cell}+(1-\eta_I)I_\mathrm{cell}U_\mathrm{cell}N_\mathrm{cell}
	\end{equation}
	\begin{equation}
	P_\mathrm{H_2}=P_\mathrm{ele}-Q_\mathrm{ele}=U_\mathrm{th}I_\mathrm{cell}N_\mathrm{cell}
	\end{equation}
\end{subequations}
in which $U_\mathrm{cell}$ is the electrolysis cell voltage and is a function of current $I_\mathrm{cell}$ and temperature $T_\mathrm{cell}$. $N_\mathrm{cell}$ is the number of cells per stack, $U_\mathrm{th}$ is \SI{1.48}{V}, and a current efficiency term $\eta_I$ is introduced in the heat production $Q_\mathrm{ele}$ calculation based on \cite{Current efficiency,3MW}.

Part of the electrolysis heat produced $Q_\mathrm{ele}$ is dissipated to the ambient environment through natural convection and radiation as $Q_\mathrm{dis,stack}$, and the remaining $Q_\mathrm{lye}$ is removed from the stack by the electrolyte. The hot electrolyte mixed with the gas product enters the gas-liquid separator, in which the gas product is separated for subsequent processing, and the remaining electrolyte streams from two sides are mixed and circulated into the stack. A cooling coil is placed in the separator to cool down the electrolyte by cooling water whose flow rate is controlled by the water valve. The heat in the electrolyte is transferred into the cooling water as $Q_\mathrm{cool}$ and dissipated to the ambient environment through forced convection by the fan in the chiller.

\subsection{Partition of thermal characteristics}
\label{sec:Partition of thermal characteristics}
The existing alkaline electrolysis system only has cooling devices without heating, which leads to different thermal characteristics in low-load and high-load regions, as shown in \reftab{tab:region}.

Here we define a thermal-neutral operation point as the division between two regions. This point corresponds to the load $P_\mathrm{th}$ or current $I_\mathrm{th}$ at which the heat produced by electrolysis $Q_\mathrm{ele}$ and the heat dissipated from the stack $Q_\mathrm{dis,stack}$ and separator $Q_\mathrm{dis,sep}$ are balanced at the temperature set point $T_\mathrm{set}$ without cooling water.
\begin{subequations}
	\begin{equation}
	Q_\mathrm{ele}(I_\mathrm{th},T_\mathrm{set})=Q_\mathrm{dis,stack}(T_\mathrm{set})+Q_\mathrm{dis,sep}(T_\mathrm{set})
	\end{equation}
	\begin{equation}
	Q_\mathrm{cool}=0
	\end{equation}
\end{subequations}

In the low-load region, the heat produced by the electrolysis reaction is smaller than the heat dissipated to the ambient; hence there is no need for further cooling. The cooling valve is closed, and the stack temperature cannot be maintained at the set point. In contrast, in the high-load region, the cooling valve is opened to cool the system. In the dynamic operation scenario, the stack temperature fluctuates around the set point whose dynamic characteristic depends on the temperature controller.

\begin{table}[htbp]
	\caption{Operation region and thermal characteristics}
	\label{tab:region}
	\centering
	\begin{adjustbox}{center}
	\begin{tabular}{cccc}
		\hline
		& Condition & Cooling valve & Stack temperature\\
		\hline
		Low-load region & $Q_\mathrm{ele}<Q_\mathrm{dis,stack}+Q_\mathrm{dis,sep}$ & Closed& Lower than the set point\\
		\tabincell{c}{Thermal-neutral\\ opreation}&$Q_\mathrm{ele}=Q_\mathrm{dis,stack}+Q_\mathrm{dis,sep}$&Closed&At set point\\
		High-load region&$Q_\mathrm{ele}>Q_\mathrm{dis,stack}+Q_\mathrm{dis,sep}$& Open&\tabincell{c}{Around the set point,\\ affected by temperature\\ controller }\\
		\hline
	\end{tabular}
\end{adjustbox}
\end{table}

This paper focuses on the thermal process in high-load region, especially the temperature overshoot during load ramping. For the lab-scale system, the thermal neutral point is high: 70\% for the \SI{25}{kW} system at an ambient temperature of 10$^{\circ}$C and a stack temperature of 70$^{\circ}$C. However, for the MW-scale system, the load of the thermal neutral point will be reduced to approximately 20\%-40\% due to the smaller proportion of heat dissipation, shown in section \ref{large-scale systems}. The high-load region becomes more general.

\subsection{Temperature control in the high-load region: trade-off between safety and efficiency}\label{s:overshoot}
\begin{figure}[htb]  
	\makebox[\textwidth][c]{\includegraphics[width=0.8\textwidth]{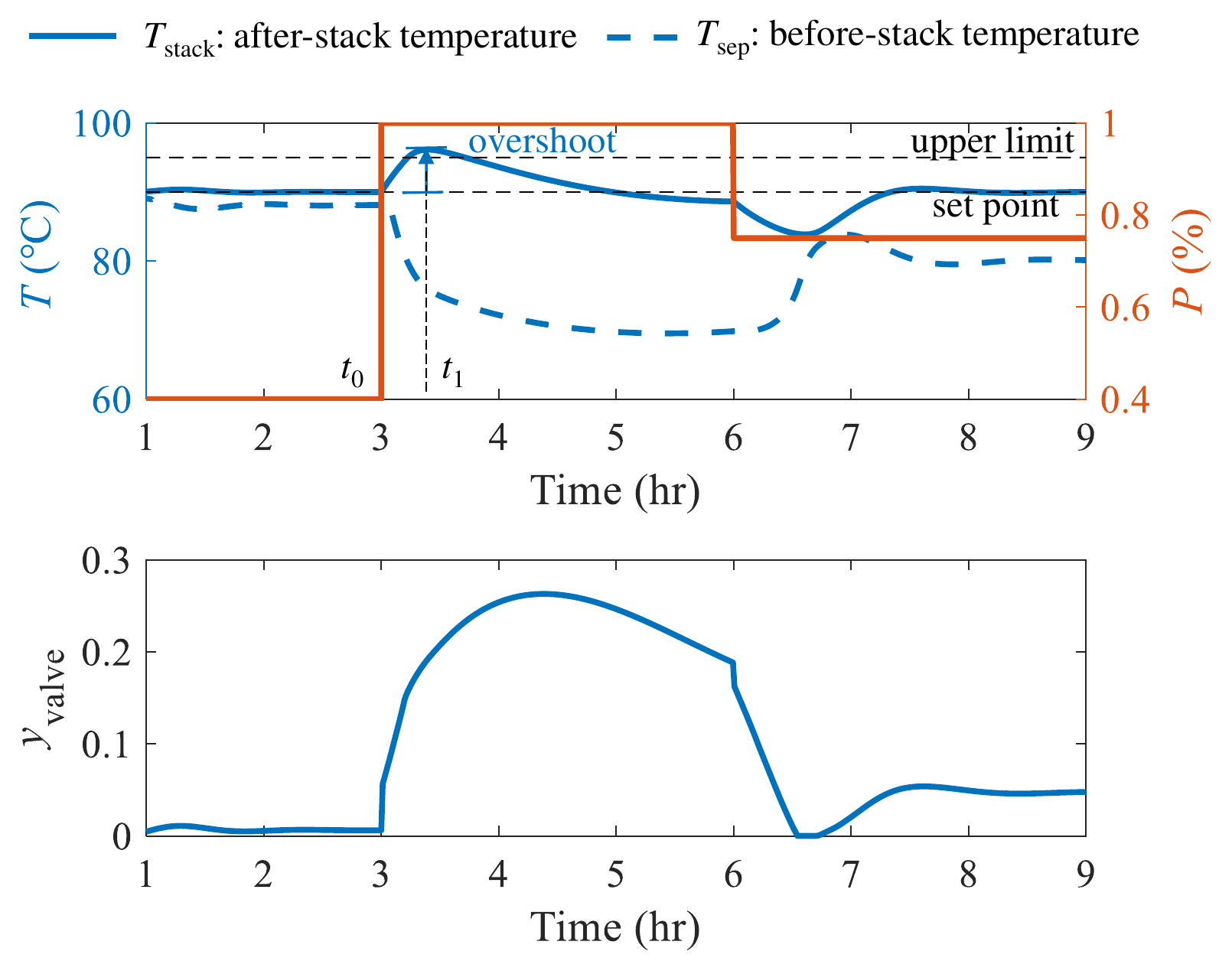}}  
	\caption{Dynamic process of the after-stack temperature with load fluctuations.}
	\label{Fig2Overshoot}
\end{figure}

In the high-load region, the system temperature cannot be fully controlled at the set point due to the large thermal inertia and time delay. As in \reffig{Fig2Overshoot}, at $t=$\SI{3}{hr}, the load $P$ ramps up which results in the after-stack temperature $T_\mathrm{stack}$ increases immediately. The after-stack temperature deviates from the set point, and the PID temperature controller opens the cooling water valve, shown as $y_\mathrm{valve}$. 
However, the cooling effect cannot be immediately applied to the after-stack temperature because it takes time from the increase in the cooling water flow rate to the decrease in the after-stack temperature $T_\mathrm{stack}$ due to the time-delay and thermal inertia, which is illustrated in \reftab{tab:Cooling process}. Furthermore, the water valve is regulated according to the temperature deviation and opens gradually rather than switched to the steady-state value directly. This slow action of the temperature controller, as well as the time-delay and thermal inertia in the thermal process results in large temperature overshoot. Though the after-stack temperature is finally back to the set point again, the stack may exceed the upper limit causing safety problems.

\begin{table}[htbp]
	\caption{The cooling process}
	\label{tab:Cooling process}
	\centering
	\begin{tabular}{|p{0.3\textwidth}<{\centering}|p{0.3\textwidth}<{\centering}|p{0.3\textwidth}<{\centering}|}
		\hline
		Time & Phenomenon & Mechanism\\
		\hline
		\multirow{3}{*}{$t_0=$\SI{3}{hr}}& Load $P$ ramping & \\
		\cline{2-3}
		&After-stack temperature $T_\mathrm{stack}$ increases&\\
		\cline{2-3}
		&Valve opening $y_\mathrm{valve}$ increases&\\
		\hline
		$t_{01}=t_0+\tau_{2}$& Before-stack temperature $T_\mathrm{sep}$ starts to decrease&Time-delay $\tau_{2}$: convection of cooling water in the coil\\
		\hline
		$t_{02}=t_0+\tau_{2}+\tau_{1}$&After-stack temperature $T_\mathrm{stack}$ is influenced by cooling&Time-delay $\tau_{1}$: convection of electrolyte in the stack\\
		\hline
		$t_1=$\SI{3.4}{hr}&After-stack temperature $T_\mathrm{stack}$ starts to decrease&$t_1-t_{02}$ is caused by the thermal inertia of the stack $C_\mathrm{stack}$\\
		\hline
	\end{tabular}
\end{table}

Considering the temperature overshoot, selecting the temperature set point becomes a challenge under dynamic operation:
\begin{itemize}
	\item High temperature set point: temperature overshoot may harm the stack.
	\item Low temperature set point: low temperature at steady-state cause efficiency loss.
\end{itemize}
To solve this problem, the following section focuses on thermal process modeling and temperature controller design. A control-oriented thermal model is established to accurately describe the temperature overshoot phenomenon. Novel temperature controllers are designed based on the feed-forward method, which opens the cooling water valve in time or in advance. By the methods above, temperature fluctuations can be suppressed and higher temperature set points are allowed to improve system efficiency.

\section{Control-oriented thermal model with multiple inertial elements and time delays}
In this section, a control-oriented thermal model is proposed based on the traditional lumped thermal model, and the novelties are summarized.
\subsection{Traditional lumped thermal model}
The most widely used thermal model was proposed by Ulleberg \cite{CiteUlleberg}:
\begin{equation}\label{eq:lumped model}
C\frac{\mathrm{d} \bar{T}}{\mathrm{d}t}=Q_\mathrm{ele}-Q_\mathrm{dis}-Q_\mathrm{cool}
\end{equation}
where $\bar{T}$ and $C$ represent the average temperature and thermal capacity of the electrolysis system, respectively. $Q_\mathrm{ele}$, $Q_\mathrm{dis}$ and $Q_\mathrm{cool}$ are the heat production by electrolysis, heat dissipated to the ambient environment and cooling demand, respectively.

This thermal model \refequ{eq:lumped model} has a simple form and is easy to use; however, it is not suitable for temperature controller design and simulation due to the lack of precision. On the one hand, it models the average system temperature $\bar{T}$ rather than the before-stack or after-stack temperature. Indeed, in high-load region, the after-stack temperature is always the hottest point in the system and is specifically controlled to within the safety region. On the other hand, the lumped model \refequ{eq:lumped model} ignores the time-delay in the thermal dynamic process and cannot accurately predict the large overshoot and oscillation.

\subsection{Proposed control-oriented thermal model}
\begin{figure}[htb]  
	\makebox[\textwidth][c]{\includegraphics[width=0.7\textwidth]{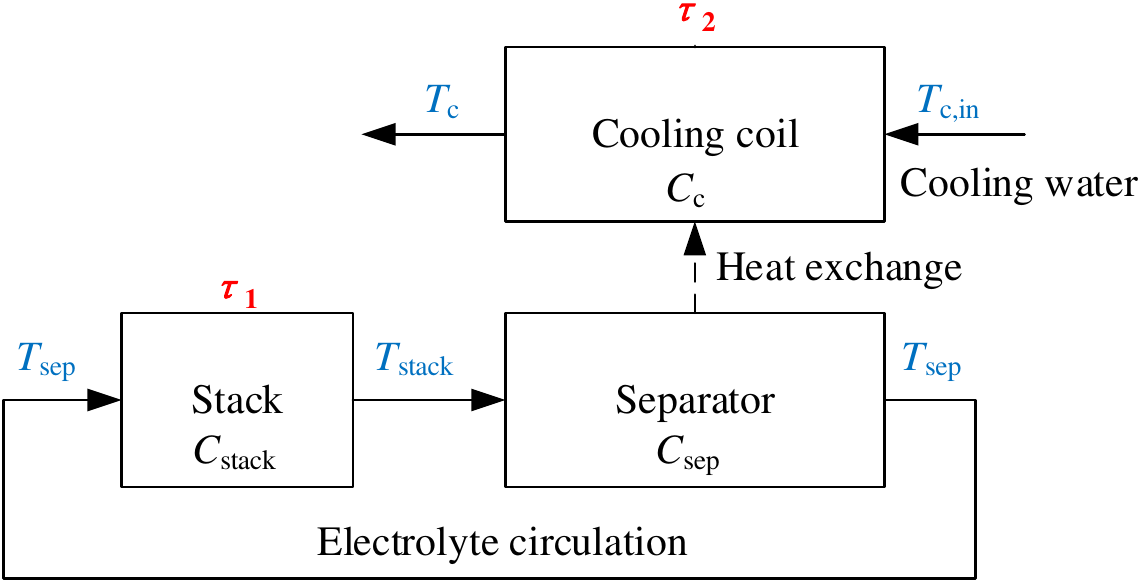}}  
	\caption{Simplified heat transfer process of the alkaline electrolysis system with key parameters.}
	\label{Fig3Modeling}
\end{figure}

Focusing on the controller design and simulation requirements, a novel model with the form of a third-order time-delay process is established in this section. The schematic diagram is shown in \reffig{Fig3Modeling}. The system's thermal inertia is abstracted into three thermal capacities $C_\mathrm{stack}$, $C_\mathrm{h}$ and $C_\mathrm{c}$, representing the stack, separator and cooling coil, respectively. Two time-delay terms $\tau_{1}$ and $\tau_{2}$ are introduced:
\begin{itemize}
	\item Stack's time-delay $\tau_{1}$: When the before-stack temperature changes, the after-stack temperature will change after a time delay $\tau_{1}$ due to electrolyte convection in the stack. 
	\item Cooling coil's time-delay $\tau_{2}$: When the flow rate of cooling water changes, the influence is delayed by $\tau_{2}$ to the temperature of the cooling water $T_\mathrm{c}$.
\end{itemize}

The thermal energy balance of the stack, separator and cooling coil can be expressed as:
\begin{subequations} \label{eq:model}
	\begin{equation}\label{eq:model_stack}
	C_{\text {stack }} \frac{\mathrm{d} T_{\mathrm{stack},t}}{\mathrm{d} t}=Q_{\text {ele },t}-Q_{\mathrm{dis,stack},t}-c_\mathrm{lye}v_\mathrm{lye}\rho_\mathrm{lye}\left(T_{\mathrm{stack},t}-T_{\mathrm{sep},t-\tau_{1}}\right) 
	\end{equation}
	\begin{equation}\label{eq:model_sep}
	C_{\mathrm{sep}} \frac{\mathrm{d} T_{\mathrm{sep},t}}{\mathrm{d} t}=\frac{1}{2}v_{\mathrm{lye}} \rho_{\mathrm{lye}} c_{ \mathrm{lye}}\left(T_{\mathrm{stack},t}-T_{\mathrm{sep},t}\right)-k A \Delta T_{t}-Q_{\mathrm{dis,sep},t}
	\end{equation}
	\begin{equation}\label{eq:model_cool}
	C_{\mathrm{c}} \frac{\mathrm{d} T_{\mathrm{c},t}}{\mathrm{d} t}=v_{\mathrm{c},t-\tau_{2}} \rho_{\mathrm{c}} c_{ \mathrm{c}}\left(T_{\mathrm{c}, \mathrm{in},t}-T_{\mathrm{c},t}\right)+k A\Delta T_{t}
	\end{equation}
\end{subequations}
where the subscript $t$ indicates the measurement time.

\refequ{eq:model_stack} shows that the change in stack temperature $T_{\mathrm{stack},t}$ depends on the heat produced by electrolysis $Q_{\mathrm{ele},t}$, the heat loss to the ambient $Q_{\mathrm{dis,stack},t}$ and the heat removed by the electrolyte $c_\mathrm{p,lye}v_\mathrm{lye}\rho_\mathrm{lye}(T_{\mathrm{stack},t}-T_{\mathrm{sep},t-\tau_{1}})$. The heat produced $Q_{\mathrm{ele},t}$ is \refequ{eq:Heatproduced}, and an empirical relationship is used for the cell voltage $U_\mathrm{cell}$ calculation:
\begin{equation}\label{eq:UIcurve}
U_{\text {cell },t}=U_{\text {rev }}+(r_{1}+r_{2} \bar{T}_t)i_t+s \log \left((t_{1}+t_{2} / \bar{T}_t+t_{3} / \bar{T}_t^{2})i_t +1\right)
\end{equation}
\begin{equation}\label{eq:average temperature}
\bar{T}_{t}=(T_{\mathrm{stack},t}+T_{\mathrm{sep},t})/2
\end{equation}
in which $i$ is the current density, $U_\mathrm{rev}$ is the reversible voltage, $\bar{T}$ is the average temperature, and $r_\mathrm{1}, r_\mathrm{2}, t_\mathrm{1}, t_\mathrm{2}, t_\mathrm{3}, s$ are parameters.
The heat loss to the ambient $Q_\mathrm{dis,stack}$ is composed of thermal convection and radiation:
\begin{equation}
Q_\mathrm{dis,stack}=Q_\mathrm{conv}+Q_\mathrm{rad}=h A_\mathrm{stack}(T_\mathrm{stack}-T_\mathrm{amb})+\sigma A_\mathrm{stack} \varepsilon_\mathrm{stack}\left(T_\mathrm{stack}^{4}-T_\mathrm{amb}^{4}\right)
\end{equation}
\begin{equation}
h=2.51 \times 0.52\left(\frac{(T_\mathrm{stack}-T_\mathrm{amb})}{\varphi_\mathrm{stack}}\right)^{0.25}
\end{equation}
where $h$ is the natural convection heat transfer coefficient obtained from well-known correlations for natural convection on horizontal cylinders \cite{2008Thermal}, $\sigma$ is the Boltzmann constant, $\varepsilon_\mathrm{stack}$ is the blackness, and $\varphi_\mathrm{stack}$ is the stack diameter. 
To calculate the enthalpy change of the electrolyte $c_\mathrm{p,lye}v_\mathrm{lye}\rho_\mathrm{lye}(T_{\mathrm{stack},t}-T_{\mathrm{sep},t-\tau_{1}})$, the time-delay $\tau_{1}$ is introduced representing the delay in the influence of the before-stack temperature $T_\mathrm{sep}$ on the after-stack temperature $T_\mathrm{stack}$.

\refequ{eq:model_sep} and \refequ{eq:model_cool} illustrate the heat exchange process from the electrolyte in the separator to the cooling water.
In \refequ{eq:model_sep}, the first term on the right-hand side is the enthalpy change of the electrolyte and $1/2$ is introduced because one side's electrolyte flow rate is half of the total flow rate. The second term calculates the heat transfer by the heat transfer coefficient $k$, area $A$ and the mean logarithmic temperature difference $\Delta T_t$, which is as follows:
\begin{equation}
\Delta T_t=\frac{\left(T_{\mathrm{stack},t}-T_{\mathrm{c},t}\right)-\left(T_{\mathrm{sep},t}-T_{\mathrm{c}, \mathrm{in},t}\right)}{\ln \left(\left(T_{\mathrm{stack},t}-T_{\mathrm{c},t}\right) /\left(T_{\mathrm{sep},t}-T_{\mathrm{c}, \mathrm{in},t}\right)\right)}.
\end{equation}
The third term is the heat dissipated to the ambient environment $Q_\mathrm{dis,sep}$. 
To estimate the heat dissipation of auxiliary devices including the separator and the pipelines, a thermal resistance $R_\mathrm{sep}$ is used to calculate $Q_\mathrm{dis,sep}$:
\begin{equation}
Q_\mathrm{dis,sep}=\frac{\bar{T}-T_\mathrm{amb}}{R_\mathrm{sep}}.
\end{equation}
In \refequ{eq:model_cool}, a time-delay term $\tau_{2}$ is introduced to show the delay from the change in cooling water flow rate $v_{\mathrm{c}}$ to temperature $T_\mathrm{c}$. The cooling water flow rate $v_{\mathrm{c}}$ is controlled by the valve opening $y_\mathrm{valve}$ as \refequ{eq:Valve}, where $k_\mathrm{valve}$ is the scale factor. 
\begin{equation}\label{eq:Valve}
v_{\mathrm{c},t}=k_\mathrm{valve}y_{\mathrm{valve},t}
\end{equation}

\section{Thermal controller design}
Temperature controllers control the stack temperature by regulating the cooing water valve. To mitigate temperature overshoot under dynamic operating conditions, two novel temperature controllers are proposed in this section including a PID controller with current feedforward (PID-I) and an MPC controller.
\begin{figure}[htb]  
	\makebox[\textwidth][c]{\includegraphics[width=0.85\textwidth]{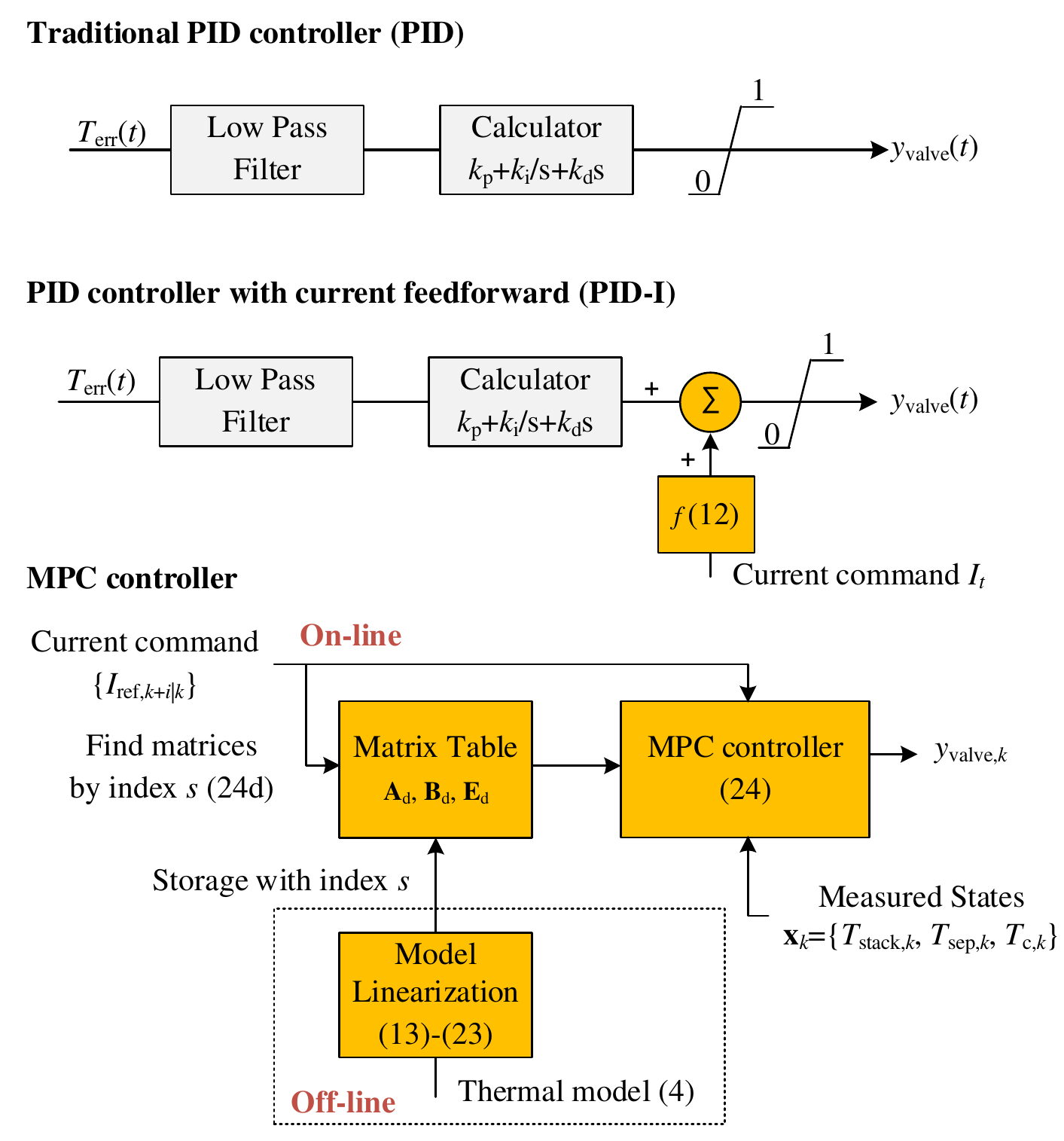}}  
	\caption{Temperature controller structures.}
	\label{Fig4Controller}
\end{figure}

\subsection{Traditional PID controller}
PID temperature controllers are the most commonly used controllers in alkaline electrolysis systems and are introduced first as a basis. The structure is shown in \reffig{Fig4Controller}, consisting of a low pass filter to remove irrelevant high frequency signals, a calculator and a limiter. To suppress supersaturation, when the output $y_\mathrm{valve}$ reaches the limit position (0 or 1), the PID integral term will no longer accumulate. The connection of the PID controller to the electrolysis system is shown in \reffig{Fig1System}(a). 

\begin{figure}[htb]  
	\makebox[\textwidth][c]{\includegraphics[width=0.8\textwidth]{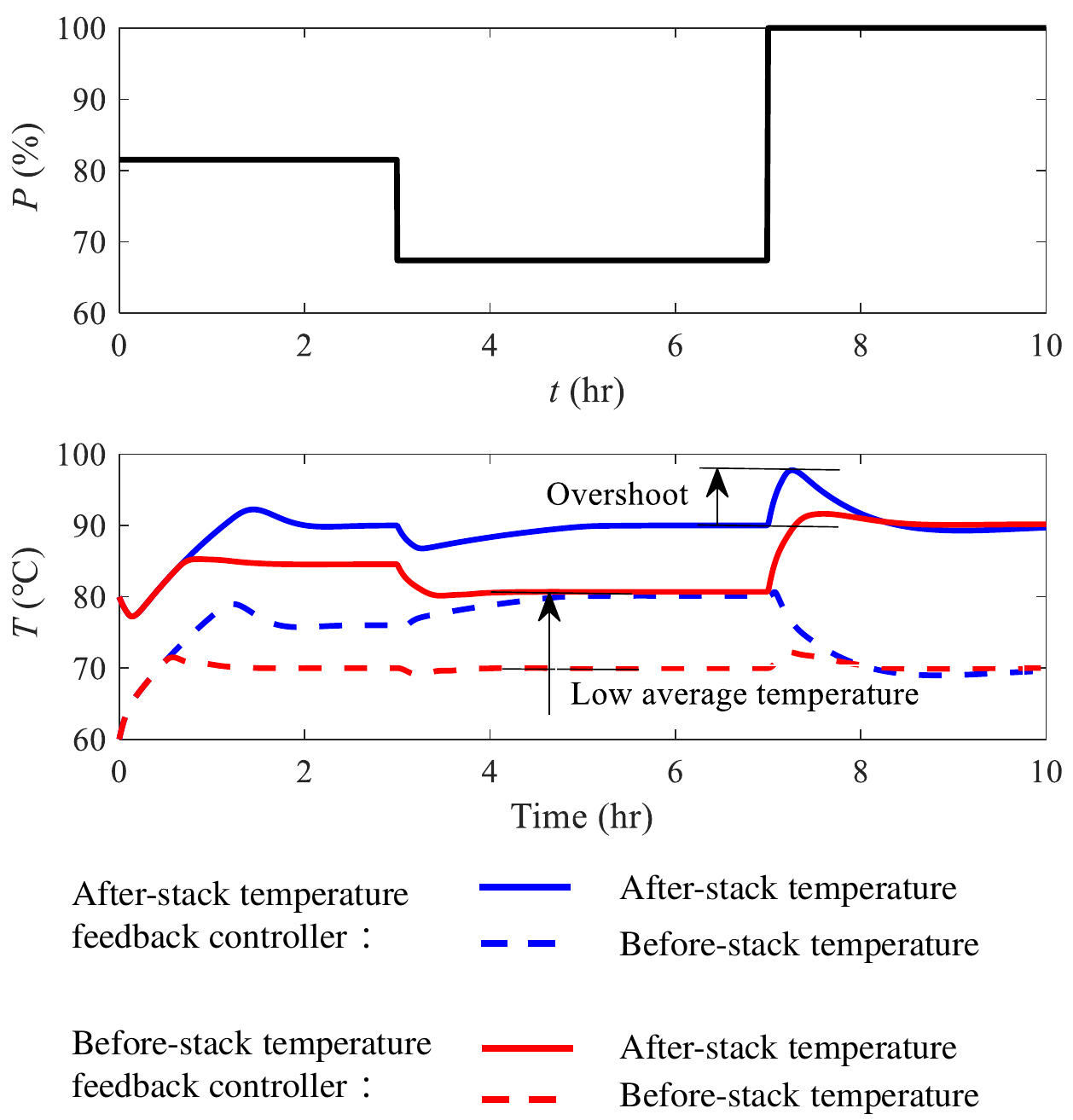}}  
	\caption{PID controller comparision: before-stack and after-stack temperature feedback}
	\label{Fig5Comparision}
\end{figure}

According to different temperature measurement points, PID controllers can be further classified into two kinds: before-stack temperature feedback controllers and after-stack temperature feedback controllers.
With the same structure as that shown in \reffig{Fig4Controller}, they have quite different performances. When the after-stack temperature is measured and controlled at the set point, the overshoot phenomenon occurs during load fluctuation which may harm the stack. The before-stack temperature feed-back controller maintains the before-stack temperature at the set point, and the after-stack temperature moves from one steady-state to another without temperature overshoot under dynamic operation; however, this results in a low average temperature of the stack leading to efficiency loss. An illustrated diagram is shown in \reffig{Fig5Comparision}, and \reftab{tab:PIDcontroller} shows the comparison.

\begin{table}[htbp]
	\caption{PID controller comparision}
	\label{tab:PIDcontroller}
	\centering
	\begin{adjustbox}{center}
		\begin{tabular}{ccc}
			\hline
			& Advantages & Disadvantages\\
			\hline
			Before-stack& Small overshoot & \tabincell{c}{Low average temperature\\ during low-load periods}\\
			After-stack& \tabincell{c}{High average temperature\\ during low-load periods} & Large overshoot\\
			\hline
		\end{tabular}
	\end{adjustbox}
\end{table}

Based on the analysis above, for an ideal temperature controller, it is preferred to choose the after-stack temperature as the controlled variable rather than the before-stack temperature considering system efficiency. Furthermore, the overshoot is expected to be reduced to ensure safe operation, which is caused by the inertia and delay in the heat transfer and convection process as illustrated in \ref{s:overshoot}.
In the following section, novel temperature controllers are proposed based on the feed-forward idea, which uses load information to regulate the water valve in time or in advance. 

\subsection{Current feed-forward PID controller (PID-I)}
The PID-I controller is designed based on the after-stack temperature feedback PID controller, shown in \reffig{Fig4Controller}. A current feed-forward term is introduced to directly give the expected valve opening value when the load changes. Compared to traditional PID controller whose output depends on the temperature error $T_\mathrm{err}$, this PID-I controller regulates the cooling water valve in a timelier manner.

The current feed-forward term is given by the linearized mapping function $f$ at a given current $I_t$ and temperature set point $T_\mathrm{set}$:
\begin{equation}\label{eq:PID-I}
f(I_t)=\frac{y_\mathrm{valve}^*(I_1,T_\mathrm{set})-y_\mathrm{valve}^*(I_2,T_\mathrm{set})}{I_1-I_2}(I_t-I_2)+y_\mathrm{valve}^*(I_2,T_\mathrm{set})
\end{equation}
in which $y_\mathrm{valve}^*(I_1,T_\mathrm{set})$ and $y_\mathrm{valve}^*(I_2,T_\mathrm{set})$ are the steady state valve openings at currents $I_1$ and $I_2$, respectively, and can be easily derived by experiments.

\subsection{Model predictive control controller}
The MPC controller adjusts the water valve opening in advance before the load changes using pre-received load information and can completely eliminate temperature fluctuations with an accurate thermal model. When the electrolysis system is used as a flexible load in the peak shaving scenario, the dispatch signal can be obtained a day in advance and is suitable for the MPC controller. Unlike the PID and PID-I controllers, the MPC controller is model-based and uses pre-received load information to achieve better performance. The structure is shown in \reffig{Fig4Controller}, which is divided into off-line and on-line parts.
\subsubsection{Off-line: linear parameter-varying model construction}
To obtain the optimal valve opening, the MPC controller solves an optimization problem online; however, a nonlinear thermal model such as \refequ{eq:model} leads to computational burden. To solve this problem, a linear parameter-varying model is used in the MPC controller design, which is derived by linearizing \refequ{eq:model} at $N_s$ steady-state operation points off-line.

The nonlinear thermal model \refequ{eq:model} can be abbreviated as \refequ{eq:non-linear thermal model}. The time-delay terms are temporarily neglected and will be introduced in the on-line part.
\begin{equation}\label{eq:non-linear thermal model}
\mathbf{\dot{x}}=\mathbf{h}(\mathbf{\dot{x}},u,I)
\end{equation}
where $\mathbf{x}$ is the state variable matrix and $u$ is the control variable:
\begin{equation}
\mathbf{x}\overset{\mathrm{def}}{=}
\left[\begin{array}{lll}
T_\mathrm{stack}& T_\mathrm{sep} & T_\mathrm{c}
\end{array}\right]^\mathrm{T}, u\overset{\mathrm{def}}{=}y_\mathrm{valve}.
\end{equation}

The first step is to make a uniform interpolation within the allowed current interval:
\begin{equation}
I(s)=\frac{(I_\mathrm{max}-I_\mathrm{min})}{N_s-1}s+I_\mathrm{min},\quad s=1,2,\cdots, N_s
\end{equation}
where $s$ is the current index in the region of $\left[I_\mathrm{min},I_\mathrm{max}\right]$ and $N_s$ is the number of steady-state points calculated. 

Then, steady-state operation points $\mathbf{x_e}^*$ are obtained at current $I(s)$ and temperature set point $T_\mathrm{set}$. $\mathbf{x_e}^*$ is defined as an expanded state vector including the  state variable $\mathbf{x^*}$, control variable $u^*$ and current $I(s)$ at steady-state:
\begin{equation}
\mathbf{x^*_e}\overset{\mathrm{def}}{=}
\left[\begin{array}{lll}
\mathbf{x^*} & u^* & I(s)
\end{array}\right]^\mathrm{T}
\end{equation}
The following equations \refequ{eq:steady-state vector} are solved for the steady-state vector $\mathbf{x_e^*}(s)$:
\begin{subequations}\label{eq:steady-state vector}
	\begin{equation}
	\mathbf{h}(\mathbf{x^*},u^*,I(s))=0
	\end{equation}
	\begin{equation}
	T_\mathrm{stack}=T_\mathrm{set}
	\end{equation}

\end{subequations}

Next, the nonlinear thermal model \refequ{eq:non-linear thermal model} is linearized at $\mathbf{x_e^*}(s)$:
\begin{equation}\label{eq:linear model}
\delta\mathbf{\dot{x}}_{t}=\mathbf{A}(s)\delta\mathbf{x}_t+\mathbf{B}(s)\delta u_t+\mathbf{E}(s)\delta I_t
\end{equation}
in which 
\begin{equation}
\delta\mathbf{x}=\mathbf{x}-\mathbf{x}^*, \delta u=u-u^*, \delta I=I-I^*
\end{equation}
and the Jacobian matrices $\mathbf{A}$, $\mathbf{B}$, $\mathbf{E}$ are as follows:
\begin{equation}
J(\mathbf{h},\mathbf{x})=\left[\begin{array}{ccc}
\frac{\partial h_{1}}{\partial x_{1}} & \cdots & \frac{\partial h_{1}}{\partial x_{n}} \\
\vdots & \ddots & \vdots \\
\frac{\partial h_{n}}{\partial x_{1}} & \cdots & \frac{\partial h_{n}}{\partial x_{n}}
\end{array}\right]_{\mathbf{x}=\mathbf{x_e^*}(s)}
\end{equation}
\begin{equation}
\mathbf{A}=J(\mathbf{h},\mathbf{x}), \mathbf{B}=J(\mathbf{h},u), \mathbf{E}=J(\mathbf{h},I).
\end{equation}

Finally, a discrete-time model can be obtained from \refequ{eq:linear model} through zero-order hold discretization:
\begin{equation}\label{discrete model}
\delta\mathbf{x}_{k+1}=\mathbf{A}_\mathrm{d}(s)\delta\mathbf{x}_k+\mathbf{B}_\mathrm{d}(s)\delta u_k+\mathbf{E}_\mathrm{d}(s)\delta I_k
\end{equation}
where 
\begin{equation}
\mathbf{A}_{\mathrm{d}}=e^{\mathbf{A} \tau_{\mathrm{s}}}, \mathbf{B}_{\mathrm{d}}=\left(\int_{0}^{\tau_{\mathrm{s}}} e^{\mathbf{A} \tau} \mathrm{d} \tau\right) \mathbf{B}.
\end{equation}
and $\tau_s$ is the sampling period. 

\subsubsection{On-line: model predictive control based on linear parameter-varying model}
At each control period $k$, we solve the following optimization problem $\mathscr{P}_k$
\begin{subequations}\label{MPC problem}
	\begin{equation}\label{MPC:objective}
	\min_{\mathbf{U}_{k}} \sum_{i=0}^{N_\mathrm{p}-1}[(\mathbf{x} _{k+i|k}-\mathbf{x}_\mathrm{set})^T\mathbf{Q}(\mathbf{x} _{k+i|k}-\mathbf{x}_\mathrm{set})+\Delta u_{k+i|i}^T\mathbf{R}\Delta u_{k+i|i}]
	\end{equation}
	\begin{equation}\label{equality constraint}
	\begin{aligned}
	\mathbf{x}_{k+i+1|k}=&\mathbf{A}_{\mathrm{d1},k+i|k}(s_{k+i|k})\mathbf{x}_{k+i|k}+\mathbf{A}_{\mathrm{d2},k+i|k}(s_{k+i|k})\mathbf{x}_{k+i-m_1|k}\\&+\mathbf{B}_{k+i|k}(s_{k+i|k})u_{k+i-m_2|i}+\mathbf{E}_{k+i|k}(s_{k+i|k})\mathbf{I}_{k+i|k}+\mathbf{e}(s_{k+i|k}),\\& i=0,1,\dots,N_p-1
	\end{aligned}
	\end{equation}
	\begin{equation}\label{inequality constraint}
	0\leq \mathbf{U}_k \leq 1
	\end{equation}
	\begin{equation}\label{index}
	s_{k+i|k}=\frac{I_{\mathrm{ref},k+i|k}-I_\mathrm{min}}{I_\mathrm{max}-I_\mathrm{min}}(N_s-1), i=0,1,\dots,N_p-1
	\end{equation}
\end{subequations}
where the subscript ${k+i|k}$ refers to the prediction value of period $k+i$ evaluated at period $k$. $N_p$ denotes the prediction horizon. The decision variable $\mathbf{U}_{k}$ is an aggregated vector of the time series of $u$ over the prediction horizon:
\begin{equation}
\mathbf{U}_{k}=\left[u_{k|k}, u_{k+1|k}, \ldots, u_{k+N_p-1|k}\right]^{\mathrm{T}}
\end{equation}

Specifically, $\mathscr{P}_k$ is composed of the following elements:
\begin{enumerate}
	\item \textbf{Input Parameters:} The input parameters are the previous controls $u$ in the time period of $[k-m_2,k-1]$, measured state $\mathbf{x}$ in $[k-m_1,k]$, and the future current command $I$ in $[k,k+N_p-1]$. $m_1$ and $m_2$ are the indexes for time-delays $\tau_1$ and $\tau_2$: 
	\begin{equation}
	m_1=\frac{\tau_1}{\tau_s}, m_2=\frac{\tau_2}{\tau_s}.
	\end{equation}
	
	\item \textbf{Cost function:} The cost function consists of two parts:
	\begin{enumerate}
		\item The penalty due to the deviation of stack temperature $T_\mathrm{stack}$ from set point $T_\mathrm{set}$:
		\begin{equation}
		(\mathbf{x} _{k+i|k}-\mathbf{x}_\mathrm{set})^T\mathbf{Q}(\mathbf{x} _{k+i|k}-\mathbf{x}_\mathrm{set})=q(T_\mathrm{stack}-T_\mathrm{set})^2.
		\end{equation}
		\item The actuator cost from valve regulation:
		\begin{equation}
		\Delta u_{k+i|i}^T\mathbf{R}\Delta u_{k+i|i}=r(u_{k+i|i}-u_{k+i-1|i})^2
		\end{equation}
	\end{enumerate}
where $q$ and $r$ are the weight coefficients.
\item \textbf{Constraints:} The following two kinds of constraints are considered:
\begin{enumerate}
	\item Equality constraints: \refequ{equality constraint} is derived from the discrete model \refequ{discrete model} to describe the thermal dynamics. To account for the time-delay effect, the matrix $\mathbf{A_\mathrm{d}}$ is split into $\mathbf{A}_{\mathrm{d1}}$ and $\mathbf{A}_{\mathrm{d2}}$:
	\begin{equation}
	\mathbf{A}_\mathrm{d1}=\left[\begin{array}{ccc}
	A_\mathrm{d}(1,1) & 0 & A_\mathrm{d}(1,3) \\
	A_\mathrm{d}(2,1) & 0 & A_\mathrm{d}(2,3) \\
	A_\mathrm{d}(3,1) & 0 & A_\mathrm{d}(3,3)
	\end{array}\right], 
	\mathbf{A}_\mathrm{d2}=\mathbf{A}_\mathrm{d}-\mathbf{A}_\mathrm{d1}.
	\end{equation}
	An $\mathbf{e}(s)$ term is introduced when removing
	the offset operator $\delta$ from \refequ{discrete model}:
	\begin{equation}
	\mathbf{e}(s)=\mathbf{x}^*(s)-(A_\mathrm{d}(s)\mathbf{x}^*(s)+B_\mathrm{d}(s)u^*(s)+E_\mathrm{d}(s)I^*).
	\end{equation}
	Note that although the coefficient matrices depend on the index parameter $s$ and vary over the prediction horizon, they are constant with given current command $I$. 
	\item Inequality constraints: The cooling water valve opening $u$ should be within $[0,1]$ as in \refequ{inequality constraint}.
\end{enumerate}
\item \textbf{Output solution:} The optimization problem $\mathscr{P}_k$ can be arranged as quadratic programming and solved by the interior point method. The formulation method of the quadratic programming problem is given in \ref{appendix1}. The optimal control variable series $\mathbf{U}_{k,\mathrm{opt}}$ is obtained whose first term is applied to the cooling water valve:
\begin{equation}
\hat{y}_{\mathrm{valve},k}=u_{k|k,\mathrm{opt}}
\end{equation}
\end{enumerate}

The procedure of the MPC controller is summarized in \reftab{tab:mpc}.
\begin{table}[htbp]
	\caption{Procedure of the MPC controller}
	\label{tab:mpc}
	\centering
	\begin{tabular}{ll}
	\hline
	Offline &\tabincell{l}{Obtain the linear discrete-time thermal model \refequ{discrete model}. Coefficient \\matrices $A_d, B_d$ and $E_d$ with index $s$ are stored in a table.}\\
	\hline
	\multirow{2}{*}{Online} & \tabincell{l}{First, use the MPC controller to calculate index $s_{k+i|k}$ by future \\current command $I_{k+i|k}$ in the prediction horizon as in \refequ{index}, \\which is
	used to search for the coefficient matrices in the table.} \\
	\cline{2-2}
	& \tabincell{l}{Then, solve the optimization problem $\mathscr{P}_k$ \refequ{MPC problem} \\to obtain the optimal valve opening $u_{k|k,\mathrm{opt}}$.}\\
	\hline
	\end{tabular}
\end{table}

\section{Model verification}
\begin{figure}[htb]  
	\makebox[\textwidth][c]{\includegraphics[width=1.4\textwidth]{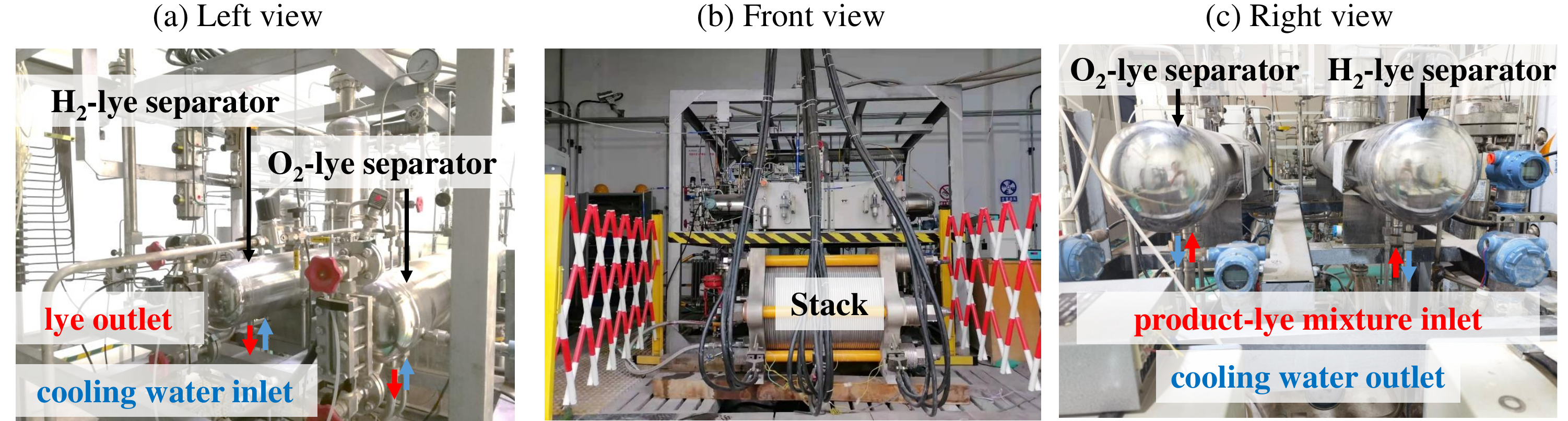}}  
	\caption{Schematic diagram of the \SI{5}{Nm^3/hr} alkaline electrolysis system CNDQ5}
	\label{Fig6Exp5N3}
\end{figure}

A commercial \SI{5}{Nm^3/hr} alkaline electrolysis system CNDQ5 from the Purification Equipment Research Institute of CSIC is used to verify the thermal model and test the controller performance, as shown in \reffig{Fig6Exp5N3}. The system process is consistent with \reffig{Fig1System}, in which a cooling coil is placed inside the gas-liquid separator of each side, and the electrolyte is cooled. The temperature of the cooling water inlet is controlled by a chiller. The upper limit of stack temperature is 95$^{\circ}$C considering the temperature tolerance of the diaphragm, however, this small-scale system cannot reach high temperatures above 85$^{\circ}$C in winter ($T_\mathrm{amb}$: 0-10$^{\circ}$C) due to relatively large surface area and heat dissipation, and the experiments are carried out at around 70-80$^{\circ}$C to verify the model.

Parameters in \ref{Parameters} are adopted for model verification. It should be clarified that the measured cell voltage $U_\mathrm{cell}$ is used rather than the UI curve because the stack performance decreases seriously during the day: cell voltage $U_\mathrm{cell}$ is the lowest after a start-up and rises slowly during daily operation, which is consistent with the phenomenon shown in \cite{Degradation}. Besides, it is found that there is still a leakage cooling water flow $v_{\mathrm{c},0}=$\SI{0.11}{m^3/hr} when the valve is closed $\hat{y}_\mathrm{valve}=0$, which is too small to be measured by the installed flow meter (measuring range 0-\SI{2}{m^3/hr}).

\begin{figure}[htb]  
	\makebox[\textwidth][c]{\includegraphics[width=0.9\textwidth]{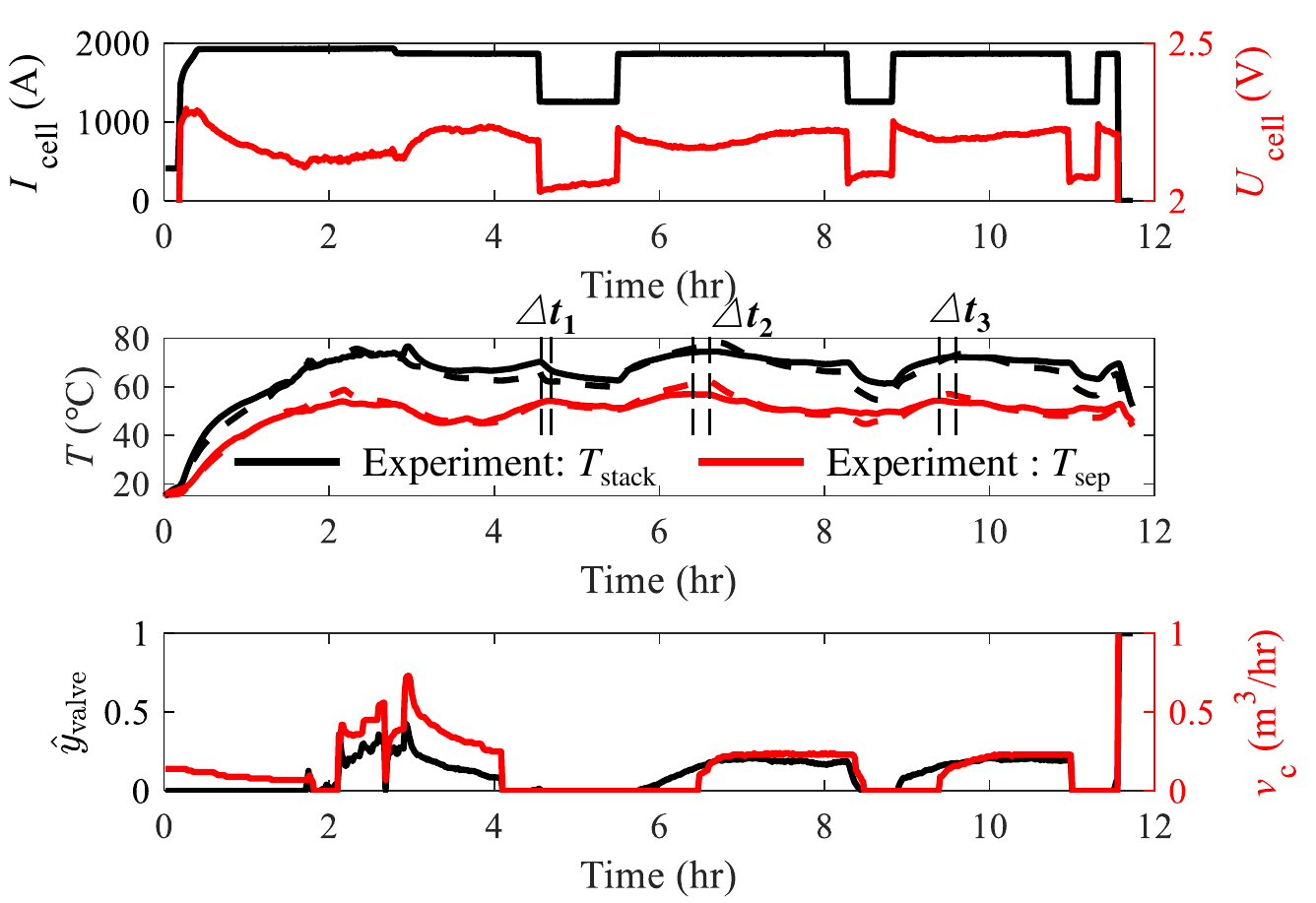}}  
	\caption{Verification of the proposed thermal model. (Full line: experimental results, Dotted line: simulation results)}
	\label{Fig7ModelVerification}
\end{figure}

\reffig{Fig7ModelVerification} shows satisfying results for the proposed thermal model \refequ{eq:model} under dynamic operating condition. Both after-stack temperature $T_\mathrm{stack}$ and before-stack temperature $T_\mathrm{sep}$ are predicted accurately, which makes finer temperature control possible, e.g., to control the after-stack temperature $T_\mathrm{stack}$ within the upper limit rather than the average temperature $\bar T$. The modelling error is caused by the linearization of the valve model shown in \reffig{FigA2Valve}. In addition, the results clearly show the time-delay effect. When the load is reduced at $t=$\SI{4.5}{hr}, the before-stack temperature $T_\mathrm{sep}$ changes $\Delta t_1$ later than the after-stack temperature $T_\mathrm{stack}$ due to the electrolyte circulation process in the separator. In contract, when the cooling valve is opened, the after-stack temperature $T_\mathrm{stack}$ changes later than the before-stack temperature $T_\mathrm{sep}$, shown as $\Delta t_2$ and $\Delta t_3$, caused by the electrolyte convection process in the stack. By introducing time-delay terms, more accurate temperature prediction can be achieved.

\section{Temperature control of small-scale systems: the experimental results}
The proposed temperature controllers are tested on the CNDQ5 experimental platform by sending control signals from an upper computer to the PLC and controlling the opening of the cooling water valve. The parameters of the temperature controllers are shown in \reftab{tab:PIDparameters}.

\begin{table}[htbp]
	\caption{Parameters of the temperature controllers}
	\centering
	\begin{tabular}{ccc}
		\hline
		&Parameters & Values \\
		\hline
		\multirow{5}{*}{PID}&Proportional coefficient $k_p$  &  20\\
		&Integral coefficient $k_i$  & 0.011 \\
		&Differential coefficient $k_d$  &6000  \\
		&Sampling period $\tau_s$& \SI{1}{s}\\
		&Time constant of the low-pass filter $T_f$  & 60s \\
		&Temperature set point $T_\mathrm{set}$  & 70$^{\circ}$C\\
		\hline
		\multirow{5}{*}{PID-I}&Steady state current $I_1$  & 720 \\
		&Steady-state current $I_2$  & 520 \\
		&Steady-state valve opening $y_\mathrm{valve}^*(I_1,T_\mathrm{set})$  & 0.11 \\
		&Steady-state valve opening $y_\mathrm{valve}^*(I_2,T_\mathrm{set})$  & 0 \\
		&Temperature set point $T_\mathrm{set}$  & 70$^{\circ}$C \\
		\hline
		\multirow{6}{*}{MPC}&Interpolation points $N_s$& 10\\
		&Prediction horizon $N_p$& 30\\
		&Sampling period $\tau_s$& \SI{2}{min}\\
		&Weight coefficient $q$& 1\\
		&Weight coefficient $r$& 300\\
		&Temperature set point $T_\mathrm{set}$  & 70$^{\circ}$C \\
		\hline
	\end{tabular}
\label{tab:PIDparameters}
\end{table}

\begin{figure}[htb]  
	\makebox[\textwidth][c]{\includegraphics[width=0.9\textwidth]{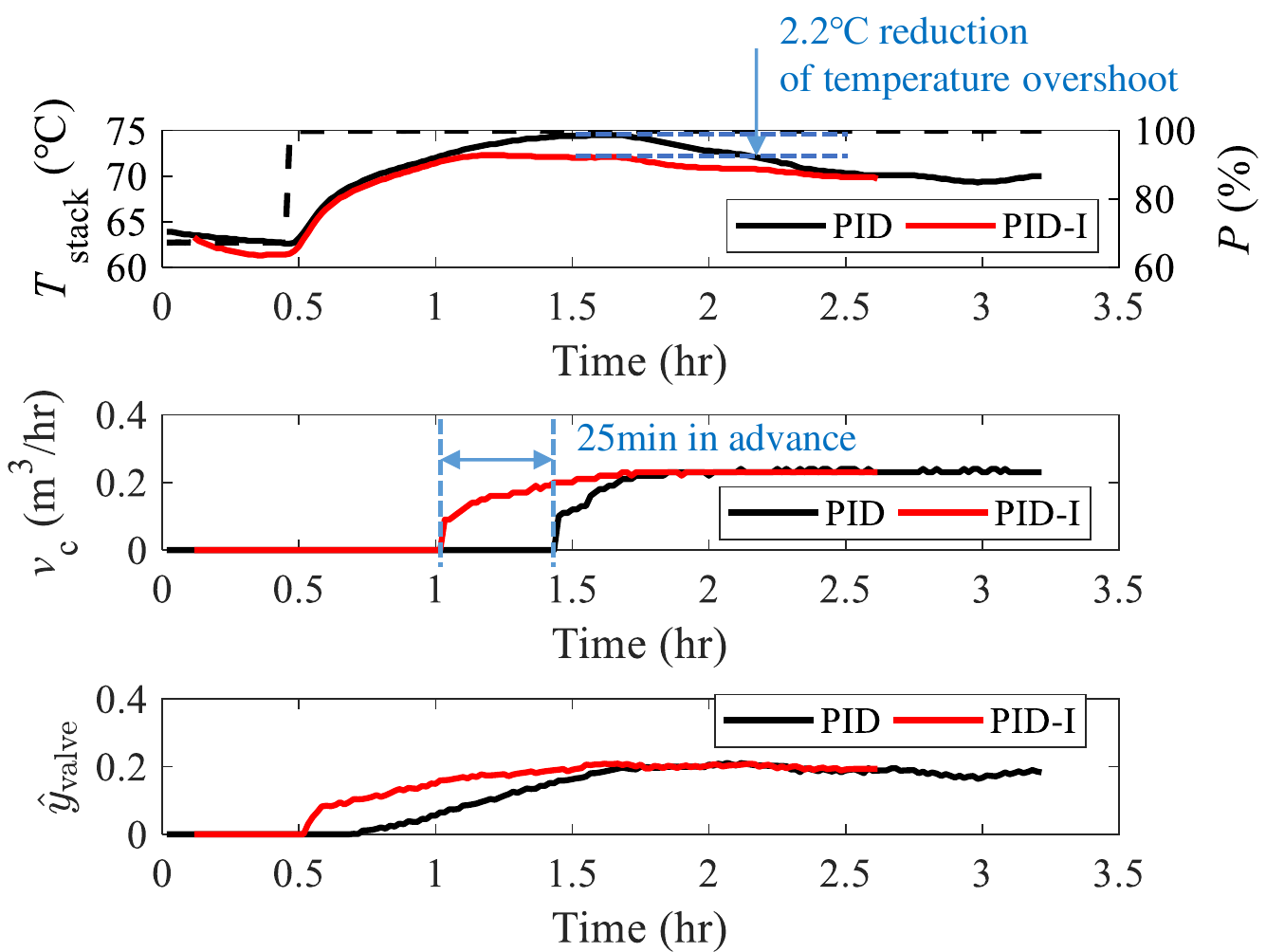}}  
	\caption{Experimental results of PID-based controllers.}
	\label{Fig8PIDcontroller}
\end{figure}

\reffig{Fig8PIDcontroller} shows the performances of PID and PID-I temperature controllers. At $t=0$, the heat produced by electrolysis $Q_\mathrm{ele}$ is insufficient to maintain the temperature set point of 70$^{\circ}$C, and the stack temperature $T_\mathrm{stack}$ reaches thermal equilibrium at approximately 67$^{\circ}$C. When the load increases from 68\% to 100\%, the increased heat production makes the after-stack temperature $T_\mathrm{stack}$ ramp up immediately. However, there is a delay in the opening of the cooling water valve for the PID controller. On the one hand, the increase in control signal $\hat{y}_\mathrm{valve}$ is a cumulative process resulting from the slow decrease in the proportional term $k_\mathrm{p}T_\mathrm{err}$; on the other hand, there is a dead zone in the valve opening process (see \reffig{FigA2Valve}), which makes the actual cooling water flow rate $v_\mathrm{c}$ start increasing at $t$=\SI{1.43}{hr}. The peak temperature of 74.5$^{\circ}$C is reached at $t$=\SI{1.58}{hr}, corresponding to a temperature overshoot of 4.5$^{\circ}$C. By contrast, the PID-I controller adds a current feed-forward term \refequ{eq:PID-I}, which boosts the valve opening $\hat{y}_\mathrm{valve}$ immediately when the load increases. By using the PID-I temperature controller, the increase in cooling water flow rate is brought forward by \SI{25}{min} resulting in the 2.2$^{\circ}$C reduction of temperature overshoot.

\begin{figure}[htb]  
	\makebox[\textwidth][c]{\includegraphics[width=0.9\textwidth]{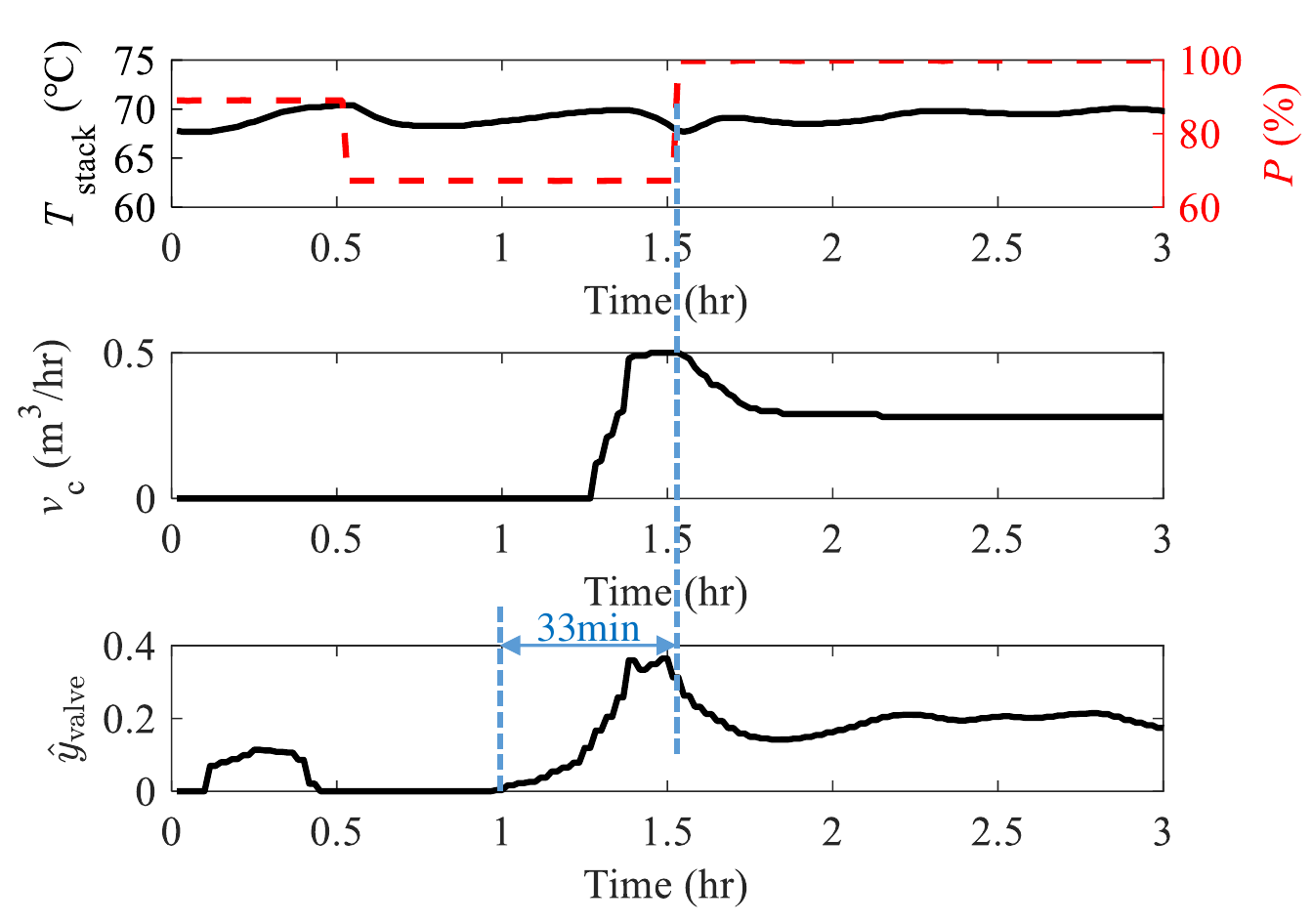}}  
	\caption{Experimental results of the MPC controller.}
	\label{Fig9MPCcontroller}
\end{figure}

The performance of the MPC controller is shown in \reffig{Fig9MPCcontroller}. The controller regulates the valve opening in advance before the load changes, because the load dispatch signal is prestored in the controller and an optimal $\hat{y}_\mathrm{valve}$ is calculated by the thermal model embedding. For example, the load $P$ increases from 68\% to 100\% at $t$=\SI{1.53}{hr}, and the control signal $\hat{y}_\mathrm{valve}$ increases \SI{33}{min} earlier which results in a stable stack temperature. The temperature overshoot phenomenon does not occur with the MPC controller.

The experimental results presented above show that both the PID-I and MPC controllers can reduce the temperature disturbance caused by load fluctuation. Compared to the traditional PID controller, the PID-I controller adds a current feed-forward term which reduces the temperature overshoot by 2.2$^{\circ}$C, and the MPC controller completely eliminates the temperature overshoot by opening the valve in advance. 

The experimental platform has a relatively small capacity of \SI{5}{Nm^3/hr}, and the experiments described above are carried out outdoors in winter with an ambient temperature $T_\mathrm{amb}$ of approximately 0-15$^{\circ}$C. This leads to a large heat loss $Q_\mathrm{dis}$ and a high thermal-neutral operation point $P_\mathrm{th}\approx70\%$. However, for large-scale systems, the heat loss $Q_\mathrm{dis}$ is much smaller than the heat produced $Q_\mathrm{ele}$, and the $P_\mathrm{th}$ can be reduced to $20-40\%$, which makes temperature control more important. The performance of thermal controllers on large-scale alkaline electrolysis systems is analyzed by simulation in the next section.

\section{Temperature control of large-scale systems}
\label{large-scale systems}
\begin{figure}[htb]  
	\makebox[\textwidth][c]{\includegraphics[width=0.9\textwidth]{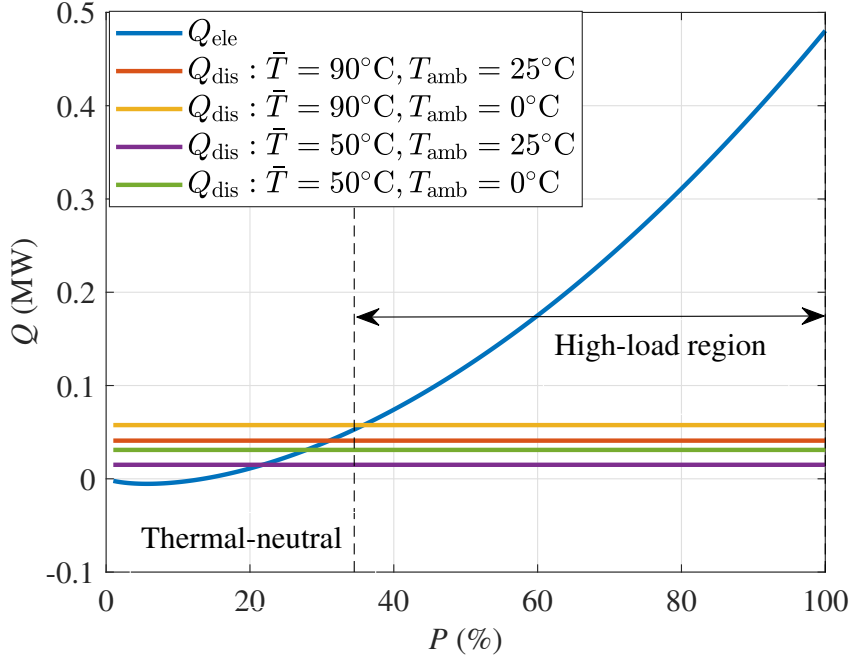}}  
	\caption{Experimental results of PID based controllers.}
	\label{Fig10Steadystate}
\end{figure}

A \SI{500}{Nm^3} alkaline electrolysis system is simulated whose parameters are derived based on the \SI{5}{Nm^3} system and engineering experience, as shown in \reftab{tab: 50AEL}. For large-scale systems, the relative surface area per hydrogen produced is small which leads to low heat dissipation to the atmosphere $Q_\mathrm{dis}$. The heat production $Q_\mathrm{ele}$ and dissipation $Q_\mathrm{dis}$ at different loading $P$ are shown in \reffig{Fig10Steadystate}. The thermal-neutral operation point is in the range of 20\%-40\%. Compared to small-scale systems, large-scale systems have much wider high-load region, and their temperature overshoot phenomenon is more general.

\begin{figure}[htb]  
	\makebox[\textwidth][c]{\includegraphics[width=1.4\textwidth]{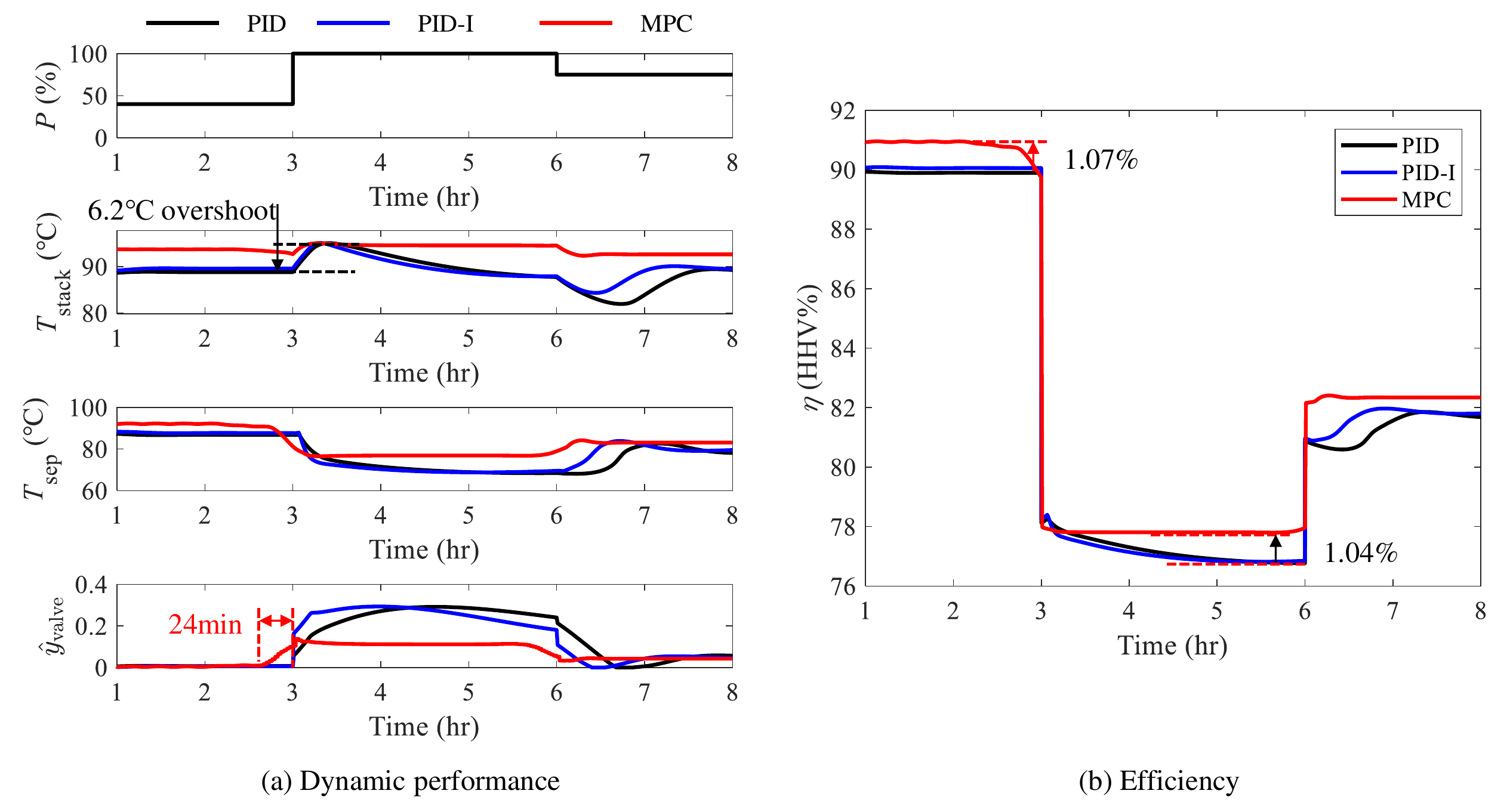}}  
	\caption{Experimental results of PID based controllers.}
	\label{Fig10Largescale}
\end{figure}

The thermal performance of a large-scale system under dynamic operating conditions is shown in \reffig{Fig10Largescale}. Under this circumstance, the temperature set point $T_\mathrm{set}$ is regulated to ensure that the after-stack temperature $T_\mathrm{stack}$ is lower than the upper limit of 95$^{\circ}$C. The PID-I controller can regulate the valve opening $\hat{y}_\mathrm{valve}$ immediately when the load ramps up. However, its effect on the temperature overshoot is not obvious in this case because we ignore the dead zone and the hysteretic characteristic of the cooling water valve in the simulation. For a real system, the dead zone of the valve (\reffig{FigA2Valve}) makes the PID-I controller more valuable, as shown in \reffig{Fig8PIDcontroller}. On the other hand, the MPC controller opens the valve \SI{24}{min} earlier before the load changes, and the temperature overshoot is reduced by 4.6$^{\circ}$C compared to that of the traditional PID controller. Therefore, a higher temperature set point $T_\mathrm{set}$ can be selected. The average stack temperature $\bar{T}$ of the MPC controller is the highest according to the after-stack temperature $T_\mathrm{stack}$ and before-stack temperature $T_\mathrm{sep}$ in \reffig{Fig10Largescale}(a). This results in an electrolysis efficiency promotion of approximately 1\% at both 40\% and 100\% loading, as shown in \reffig{Fig10Largescale}(b). By using novel temperature controllers, the disturbance effect of dynamic operation on stack temperature is weakened, which increases the electrolysis efficiency by making a higher temperature set point available and is also beneficial to prolong the system lifetime.

\begin{table}[htbp]
	\caption{Temperature set point $T_\mathrm{set}$ for different controllers.}
	\centering
	\begin{tabular}{cccc}
		\hline
		&PID & PID-I&MPC \\
		\hline
		$T_\mathrm{set}$&88.78$^{\circ}$C &89.57$^{\circ}$C &93.23$^{\circ}$C\\
		\hline
	\end{tabular}
	\label{tab:Temperature set point}
\end{table}

\section{Conclusion}
In an alkaline electrolysis system, the stack temperature is disturbed by load changes under dynamic operating conditions. In particular, temperature overshoot occurs when the load ramps up immediately which can exceed the upper limit and harm the stack.

This paper proposes novel temperature controllers to reduce the temperature overshoot and realize stable temperature control. As a basis, a control-oriented thermal model is established in the form of a third-order time-delay process, which is used for simulation and controller design. Two temperature controllers are proposed including a PID-I controller and an MPC controller, whose performances are tested on a commercial \SI{5}{Nm^3/hr} alkaline electrolysis system. The experimental results show a 2.2$^{\circ}$C overshoot reduction by the PID-I controller, and no obvious overshoot is observed by the MPC controller.
Large-scale electrolysis systems have a smaller relative heat dissipation which makes the temperature overshoot phenomenon more general. The performance of a \SI{500}{Nm^3/hr} system is analyzed by simulation. Due to the reduced temperature overshoot, the MPC controller can achieve a higher temperature set point, which increases the efficiency by approximately 1\%. The proposed temperature controllers are compared in \reftab{tab:Controllers}.

\begin{table}[htbp]
	\caption{Comparison of the proposed temperature controllers}
	\centering
	\begin{adjustbox}{center}
	\begin{tabular}{ccc}
		\hline
		&PID-I & MPC \\
		\hline
		Type&Model-free&Model-based\\
		Advantages&Easy to realize &  \tabincell{c}{Much smaller temperature overshoot,\\high efficiency and safe operation.}\\
		Disadvantages& Limited effects   & \tabincell{c}{Complex design, need the \\load information in advance} \\
		\hline
	\end{tabular}
\end{adjustbox}
	\label{tab:Controllers}
\end{table}

Although this study focuses on alkaline electrolysis systems, the temperature fluctuation under dynamic operating conditions is a common problem for both alkaline and PEM (proton exchange membrane) electrolysis systems, and the control strategy proposed in this paper is suitable for different system structures.

\section*{Acknowledgement}
This work was financially supported by the National Key R\&D Program of China (2021YFB4000500).

\appendix
\section{Quadratic programming problem formulation for the MPC controller} 
\label{appendix1}
The optimization problem \refequ{MPC problem} can be formulated as a quadratic programming problem and solved by the interior point method. The procedure is as follows.

For the convenience of the subsequent process, we expand the state variable to include previous states:
\begin{equation}
\mathbf{x}_{k}^{'}\overset{\mathrm{def}}{=}
\left[\begin{array}{llll}
\mathbf{x}_{k-m_2}^\mathrm{T}& \mathbf{x}_{k-m_2+1}^\mathrm{T} & \ldots & \mathbf{x}_{k}^\mathrm{T}
\end{array}\right]^\mathrm{T}
\end{equation}
The equality constraint \refequ{equality constraint} can be rearranged as follows:
\begin{equation}\label{k+1}
\mathbf{x'}_{k+1|k}=\mathbf{A'}(s_{k|k})\mathbf{x'}_{k}+\mathbf{B'}(s_{k|k})u_{k-m_2}+\mathbf{E'}(s_{k|k})I_k+\mathbf{e}_{k|k}
\end{equation}
where
\begin{equation}
\mathbf{A'}=\left[\begin{array}{cccc}
0 & \mathbf{I} & \cdots & 0 \\
\vdots &\vdots &\ddots &\vdots\\
0&0&\cdots&\mathbf{I}\\
\mathbf{A}_\mathrm{d2}& 0 &\cdots & \mathbf{A}_\mathrm{d1}
\end{array}\right], 
\mathbf{B'}=\left[\begin{array}{c}
0\\
\vdots\\
0\\
\mathbf{B}_\mathrm{d}
\end{array}\right], 
\mathbf{E'}=\left[\begin{array}{c}
0\\
\vdots\\
0\\
\mathbf{E}_\mathrm{d}
\end{array}\right], 
\end{equation}
The state of period $k+2$ evaluated at period $k$ is calculated by recursion:
\begin{equation}
\begin{aligned}
\mathbf{x'}_{k+2|k}=&\mathbf{A'}(s_{k+1|k})(\mathbf{A'}(s_{k|k})\mathbf{x'}_{k}+\mathbf{B'}(s_{k|k})u_{k-m_2}+\mathbf{E'}(s_{k|k})I_k+\mathbf{e}_{k|k})\\&+\mathbf{B'}(s_{k+1|k})u_{k-m_2+1}+\mathbf{E'}(s_{k+1|k})I_{k+1|k}+\mathbf{e}_{k+1|k}
\end{aligned}
\end{equation}
and the state of period $k+i$ evaluated at period $k$ can be concluded:
\begin{equation}\label{k+i}
\begin{aligned}
\mathbf{x'}_{k+i|k}=&\prod_{j=0}^i\mathbf{A'}(s_{k+j|k})\mathbf{x'}_{k}+\sum_{j=1}^{i}\prod_{l=0}^{i-j}\mathbf{A'}(s_{k+l|k})\mathbf{B'}(s_{k+j-1|k})u_{k+j-1-m_2|k}\\&+\sum_{j=1}^{i}\prod_{l=0}^{i-j}\mathbf{A'}(s_{k+l|k})\mathbf{E'}(s_{k+j-1|k})I_{k+j-1|k}+\sum_{j=1}^{i}\prod_{l=0}^{i-j}\mathbf{A'}(s_{k+l|k})\mathbf{e}_{k+j-1|k}
\end{aligned}
\end{equation}

We define the future state matrix $\mathbf{X}_k$ as follows:
\begin{equation}
\mathbf{X}_k\overset{\mathrm{def}}{=}
\left[\begin{array}{llll}
\mathbf{x}_{k+1|k}^\mathrm{T}& \mathbf{x}_{k+2|k}^\mathrm{T} & \ldots & \mathbf{x}_{k+N|k}^\mathrm{T}
\end{array}\right]^\mathrm{T}
\end{equation}
Then, the future states can be predicted using \refequ{k+1}-\refequ{k+i}:
\begin{equation}\label{future states}
\mathbf{X}_{k}=\mathbf{\Phi}_k \mathbf{x}_{k}^{\prime}+\mathbf{\Theta}_{1,k} \mathbf{u}_{k}^{\prime}+\mathbf{\Theta}_{2,k} \mathbf{U}_{k}+
\mathbf{\Omega} _k\mathbf{I}_{k}+\mathbf{\Gamma}_k \mathbf{e}_{k}
\end{equation}
where
\begin{equation}
\mathbf{U}_{k}\overset{\mathrm{def}}{=}\left[u_{k|k}, u_{k+1|k}, \ldots, u_{k+N_p-1|k}\right]^{\mathrm{T}}
\end{equation}
\begin{equation}
\mathbf{u}_{k}^{\prime}\overset{\mathrm{def}}{=}
\left[\begin{array}{llll}
u_{k-m_1}& u_{k-m_1+1} & \ldots & u_{k-1}
\end{array}\right]^\mathrm{T}
\end{equation}
\begin{equation}
\mathbf{I}_{k}\overset{\mathrm{def}}{=}\left[\begin{array}{llll}
I_{k}& I_{k+1} & \ldots & I_{k+N_p-1}
\end{array}\right]^\mathrm{T}
\end{equation}

The objective of the optimization problem \refequ{MPC problem} can be formulated in a matrix form:
\begin{equation}
\min_{\mathbf{U}_{k}}\mathbf{J}_k
\end{equation}
\begin{equation}\label{objective in matrix form}
\mathbf{J}_k=(\mathbf{X}_k-\mathbf{X}_\mathrm{set})^\mathrm{T}\mathbf{Q}(\mathbf{X}_k-\mathbf{X}_\mathrm{set})+(\mathbf{M}\mathbf{U}_k-\mathbf{N}_k)^\mathrm{T}\mathbf{R}(\mathbf{M}\mathbf{U}_k-\mathbf{N}_k)
\end{equation}
where
\begin{equation}
\mathbf{M}=\left[\begin{array}{ccccc}
1 &  &  &  & \\
-1 & 1 & & & \\
 & \ddots & \ddots & & \\
 & & & 1 & \\
 & & & -1& 1
\end{array}\right], 
\mathbf{N}_k=\left[\begin{array}{c}
u_{k-1}  \\
0\\
\vdots\\
0 \\
0
\end{array}\right].
\end{equation}

By substituting \refequ{future states} into \refequ{objective in matrix form}, the final quadratic programming problem is derived:
\begin{equation}
\min_{\mathbf{U}_{k}} \frac{1}{2}\mathbf{U}_{k}^\mathrm{T}\mathbf{H}\mathbf{U}_{k}+\mathbf{f}^\mathrm{T}\mathbf{U}_{k}
\end{equation}
\begin{equation}
\mathbf{H}=2(\mathbf{\Theta}_{2,k}^\mathrm{T}\mathbf{Q}\mathbf{\Theta}_{2,k}+\mathbf{M}^\mathrm{T}\mathbf{R}\mathbf{M})
\end{equation}
\begin{equation}
\mathbf{f}=2((\mathbf{\Phi}_k \mathbf{x}_{k}^{\prime}+\mathbf{\Theta}_{1,k} \mathbf{u}_{k}^{\prime}+\mathbf{\Omega} _k\mathbf{I}_{k}+\mathbf{\Gamma}_k \mathbf{e}_{k}-\mathbf{X}_\mathrm{set})^\mathrm{T}\mathbf{Q}\mathbf{\Theta}_{2,k}-\mathbf{N}_k^\mathrm{T}\mathbf{R}\mathbf{M}).
\end{equation}

\section{Parameters for alkaline electrolysis systems}
\label{Parameters}
\subsection{Thermal model parameters}
Thermally related parameters for the \SI{5}{m^3/hr} and \SI{500}{Nm^3/hr} alkaline electrolysis systems are shown in \reftab{tab: 50AEL}.

\begin{table}[htbp]
	\caption{Parameters for the alkaline electrolysis systems}
	\centering
	\begin{threeparttable}
	\begin{tabular}{ccc}
		\hline
		Parameters & \SI{5}{Nm^3/hr}&\SI{500}{Nm^3/hr}\\
		\hline
		Cell number $N_\mathrm{cell}$& 26&298\\
		Cell diameter & \SI{0.5}{m}&\SI{1.6}{m}\\ 
		Cell area $A_\mathrm{cell}$ & \SI{0.196}{m^2}&\SI{2}{m^2}\\
		Stack diameter $\varphi_\mathrm{stack}$ & \SI{0.61}{m}&\SI{2.04}{m}\\
		Stack length $L_\mathrm{stack}$& \SI{0.267}{m}&\SI{5.4}{m}\\
		Stack surface area $A_\mathrm{stack}$& \SI{1.1}{m^2}&\SI{41}{m^2}\\
		Electrode volume $V_\mathrm{stack,electrode}$ & \SI{0.03}{m^3}&-\\
		Free stack volume $V_\mathrm{stack,free}$& \SI{0.05}{m^3}&\SI{8}{m^3}\\
		Stack void fraction at rated $f_\mathrm{v}$ &0.5&0.5\\
		Current efficiency $\eta_I$ $^*$  & 1 & 1 \\
		Blackness of the stack surface $\varepsilon_\mathrm{stack}$ & 0.8&0.8\\
		Separator diameter $\varphi_\mathrm{sep}$ & \SI{0.219}{m}&-\\
		Separator length $L_\mathrm{sep}$ & \SI{2}{m}&-\\
		Separator volume $V_\mathrm{sep}$ & \SI{1.38}{m^3}&\SI{2.2}{m^3}\\
		Separator liquid level $h_\mathrm{l,sep}$& 50\% &50\%\\
		Lye composition & KOH&KOH\\
		Mass fraction of KOH in electrolyte $w_\mathrm{lye}$ & 31.2\%&31.2\%\\
		Thermal resistance $R_\mathrm{sep}$ & \SI{0.04}{K/W}&\SI{0.004}{K/W}\\
		Stack heat capacity $C_\mathrm{stack}$& \SI{120}{KJ/K}&\SI{55}{MJ/K}\\
		Separator heat capacity $C_\mathrm{sep}$ &\SI{146}{KJ/K}&\SI{4.26}{MJ/K}\\
		Cooling coil heat capacity $C_\mathrm{c}$&\SI{23}{KJ/K}&\SI{1.15}{MJ/K}\\
		Stack time-delay $\tau_{1}$ & \SI{6}{min}&\SI{6}{min}\\
		Cooling coil time-delay $\tau_{2}$ & \SI{4}{min}&\SI{4}{min}\\
		\hline
	\end{tabular}
\begin{tablenotes}
	\item[*] The current efficiency $\eta_I$ is taken as an undetermined parameter and adjusted according to experimental results.
\end{tablenotes}
\end{threeparttable}
\label{tab: 50AEL}
\end{table}

The heat capacities $C_\mathrm{stack}$, $C_\mathrm{sep}$ and $C_\mathrm{c}$ are calculated as follows:
\begin{equation}
C_\mathrm{stack}=C_\mathrm{electrode}+C_\mathrm{lye}=V_\mathrm{stack,electrode}\rho_\mathrm{steel}c_\mathrm{p,steel}+f_\mathrm{v}V_\mathrm{stack,free}\rho_\mathrm{lye}c_\mathrm{p,lye}
\end{equation}
\begin{equation}
C_\mathrm{sep}=h_\mathrm{l,sep}V_\mathrm{sep}\rho_\mathrm{lye}c_\mathrm{p,lye}
\end{equation}
\begin{equation}
C_\mathrm{c}=V_\mathrm{cool}\rho_\mathrm{H_2O}c_\mathrm{p,H_2O}
\end{equation}

\subsection{Cell U-I curve}
\begin{figure}[htb]  
	\makebox[\textwidth][c]{\includegraphics[width=0.7\textwidth]{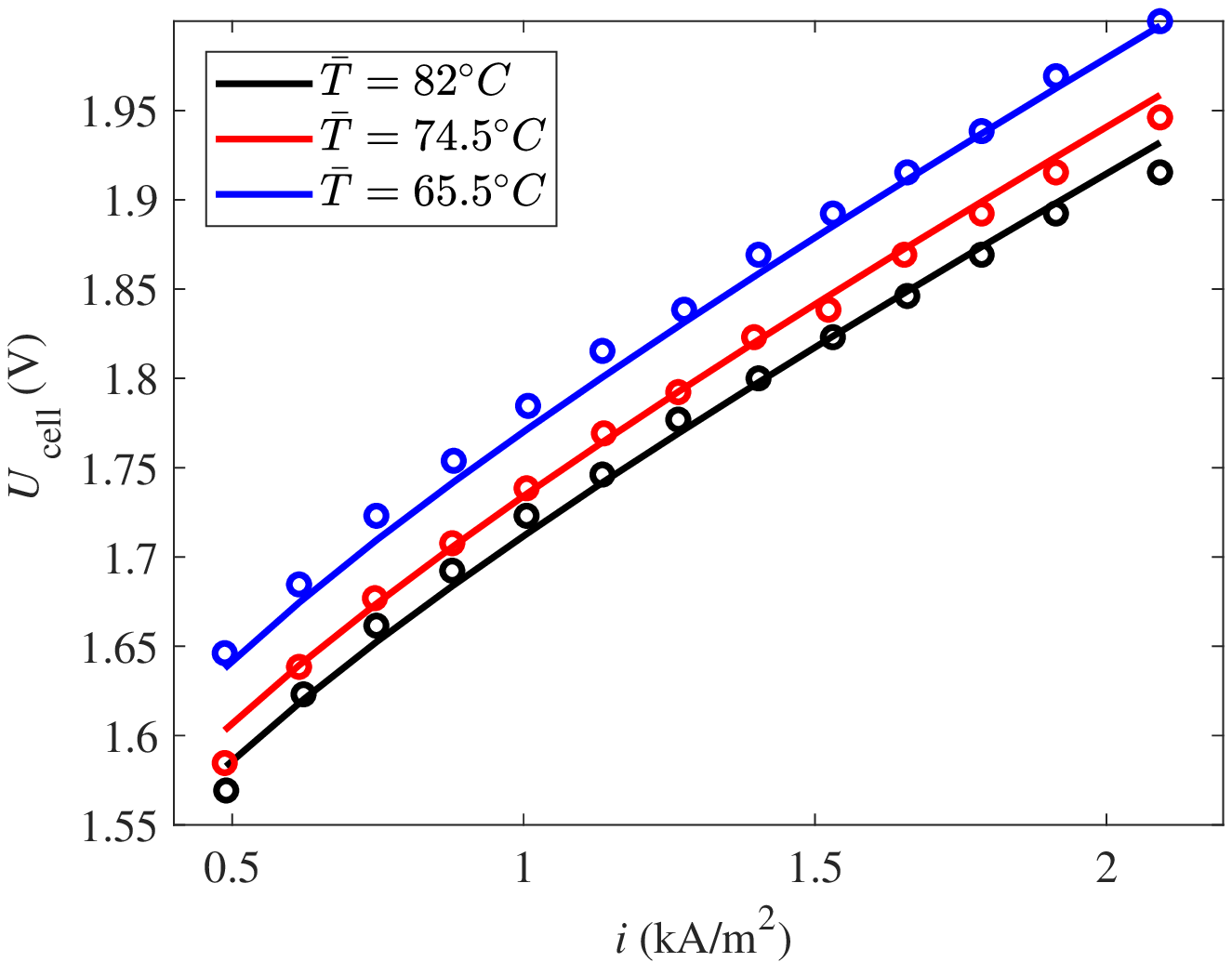}}  
	\caption{U-I curve. (o: Measured, -: Predicted.)}
	\label{FigA1UI}
\end{figure}

The U-I curve of the cell \refequ{eq:UIcurve} is determined by the curve fitting method in \cite{M-SD-experience-1-Ulleberg}, and the parameters are shown in \reftab{tab:UIcurve}. A comparison between the measured and predicted U-I curves is shown in \reffig{FigA1UI}.

\begin{table}[htbp]
	\caption{U-I curve parameters}
	\centering
	\begin{tabular}{cc}
		\hline
		Parameters & Values \\
		\hline
		$r_1$  & \SI{1.71e-4}{m^2} \\
		$r_2$ & \SI{-1.96e-7}{m^2/K}\\ 
		$s$ & \SI{0.16}{V}\\
		$t_1$ & \SI{-0.24}{m^2/A}\\
		$t_2$ & \SI{26.23}{m^2K/A}\\
		$t_3$ & \SI{139.88}{m^2K^2/A}\\
		\hline
	\end{tabular}
\label{tab:UIcurve}
\end{table}

\subsection{Valve characteristics}
\begin{figure}[htb]  
	\makebox[\textwidth][c]{\includegraphics[width=0.7\textwidth]{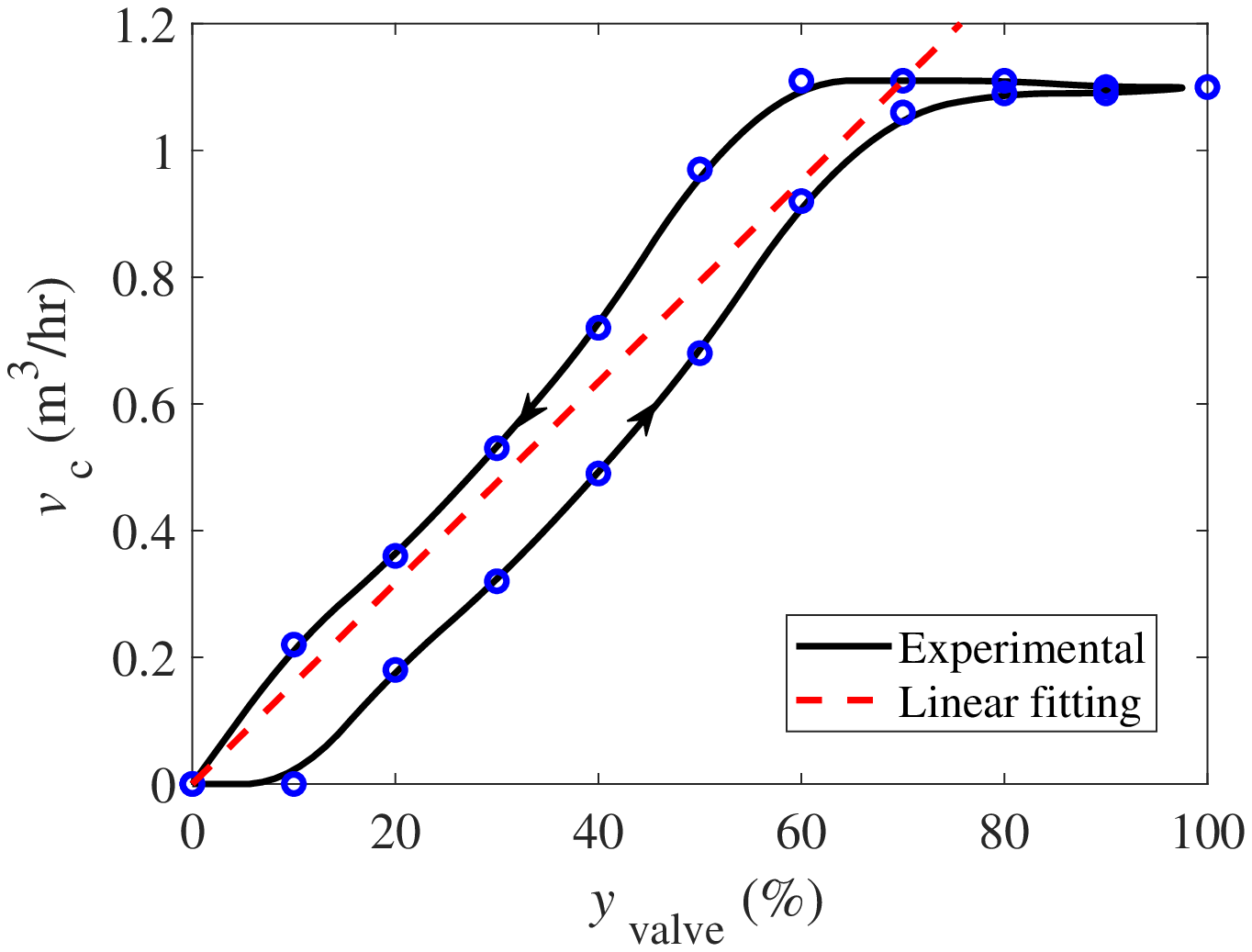}}  
	\caption{Characteristic curve of the cooling water valve.}
	\label{FigA2Valve}
\end{figure}

The cooling water valve has a hysteretic characteristic as shown in \reffig{FigA2Valve}: in the opening and closing process, the same command signal $y_\mathrm{valve}$ corresponds to different flow rates $v_\mathrm{c}$. In this paper, the model nonlinearity due to hysteresis is ignored, and a linear relationship \refequ{eq:Valve} is adopted to fit the valve characteristics. The scale factor $k_\mathrm{valve}$ in \refequ{eq:Valve} is \SI{1.1}{m^3/hr}. 

\end{document}